%% file: bare_jrnl.tex
\documentclass[journal]{IEEEtran}

\usepackage{float, siunitx}
\usepackage{soul, color, acronym, multirow, amsmath, textcomp, balance, subcaption}
\usepackage{url, tabularx}
\usepackage[justification=centering]{caption}
\usepackage[font={small}]{caption}
\usepackage[export]{adjustbox}
\usepackage{graphicx}

\input{acr.tex}

\begin{document}
\bstctlcite{IEEEexample:BSTcontrol}
\title{Environmental Monitoring for Smart Cities}

\author{
	 \IEEEauthorblockN{Bacco Manlio, Delmastro Franca, Ferro Erina, Gotta Alberto}

    \thanks{Bacco Manlio, Ferro Erina and Alberto Gotta are researchers of the Institute of Information Science and Technologies, National Research Council (CNR), Via Moruzzi 1, 56124 Pisa (Italy)}
    
    \thanks{Delmastro Franca is a researcher of the Institute of Informatics and Telematics, National Research Council (CNR), Via Moruzzi 1, 56124 Pisa (Italy)}
    
    \thanks{Corresponding author: Bacco Manlio, e-mail: manlio.bacco@isti.cnr.it}
    
    \thanks{ \copyright 2017 IEEE.  Personal use of this material is permitted.  Permission from IEEE must be obtained for all other uses, in any current or future media, including reprinting/republishing this material for advertising or promotional purposes, creating new collective works, for resale or redistribution to servers or lists, or reuse of any copyrighted component of this work in other works.}      
}

\maketitle

\begin{abstract}
This work presents an innovative, multidisciplinary and cost-effective ecosystem of ICT solutions able to collect, process and distribute geo-referenced information about the influence of pollution and micro-climatic conditions on the quality of life in Smart Cities. The system has been developed and experimentally evaluated in the framework of the research project SHE, co-funded by the Tuscany Region (Italy). Specifically, an innovative monitoring network has been developed, constituted by fixed and mobile sensor nodes, which provided comparable measurements in stationary and mobile conditions. In addition, sensor data have been enriched with those generated by citizens through the use of a dedicated mobile application, exploiting {\it participatory sensing} and {\it MSN} paradigms. 
\end{abstract}


\IEEEpeerreviewmaketitle

\input{intro-FD}
\input{related-FD}
\input{scenario-FD}
\input{results}
\input{conclusions}


\section*{Acknowledgments}
The authors would like to thank 
Pisa's municipality, PISAMO, the project coordinator Infomobility srl, 
Digitech srl, Rigel Engineering srl, UNIFI-DISPAA, IFC-CNR, CNR-IIT, CNR-ISTI. 
The SHE project has been co-funded by Tuscany Region within the \textit{Call R\&D 2012} of the POR CREO FESR 2007-2013 regional programme.


\balance
\bibliographystyle{IEEEtran}
\bibliography{bare_jrnl}
\vskip -1\baselineskip plus -1fil
\begin{IEEEbiographynophoto}{Manlio Bacco}
is a researcher at the Institute of Science and Information Technologies (ISTI) of the Italian National Research Council (CNR) in Pisa. He holds a Ph.D. in Information Engineering and Science from the University of Siena. His research interests include wireless communications, with a focus on satellite and aerial networks, and \ac{IoT}/\ac{M2M} communications. He participated and is still part of European-funded, national-funded and \ac{ESA}-funded research and development projects.
\end{IEEEbiographynophoto}
\vskip -2\baselineskip plus -1fil
\begin{IEEEbiographynophoto}{Franca Delmastro}
is a researcher of the Institute of Informatics and Telematics (IIT) of the Italian National Research Council (CNR) in Pisa since 2007. She received the Master degree in Computer Engineering and the PhD in Information Engineering, both from the University of Pisa, in 2002 and 2006 respectively. Her research interests are in the area of pervasive and ubiquitous computing including opportunistic and sensor networks, with particular attention to middleware solutions for context- and social-aware applications for Smart Cities and Health and Well-being scenarios.
\end{IEEEbiographynophoto}
\vskip -2\baselineskip plus -1fil
\begin{IEEEbiographynophoto}{Erina Ferro}
received her Laurea degree with distinction in Computer Science from the University of Pisa (Italy) in 1975. Since 1976 she is with CNR-ISTI as a Director of Research at the Wireless Networks research Laboratory.
Her main research activities are in wireless communications, especially sensor networks applied to cultural heritage, health\&well being, and AAL.   
\end{IEEEbiographynophoto}
\vskip -2\baselineskip plus -1fil
\begin{IEEEbiographynophoto}{Alberto Gotta}
(University of Genoa, MS’02-PhD’07) is a researcher at the Wireless Networks Laboratory at CNR-ISTI, Italy. His expertise is mainly related to architectures for terrestrial wireless and satellite networks applied in the context of ubiquitous networks for multimedia traffic and environmental monitoring.
\end{IEEEbiographynophoto}


\end{document}

%% file: acr.tex
\acrodef{PER}{Packet Erasure Rate}
\acrodef{PLR}{Packet Loss Rate}
\acrodef{BER}{Bit Error Rate}
\acrodef{RTT}{Round-Trip Time}
\acrodef{TCP}{Transmission Control Protocol}
\acrodef{AIMD}{Additive Increase Multiplicative Decrease}
\acrodef{PEP}{Performance Enhancing Proxy}
\acrodef{cwnd}{congestion window}
\acrodef{STP}{Satellite Transport Protocol}
\acrodef{BDP}{Bandwidth-Delay Product}
\acrodef{QoS}{Quality of Service}
\acrodef{LoS}{Line of Sight}
\acrodef{NLoS}{Non Line of Sight}
\acrodef{BLoS}{Beyond Line of Sight}
\acrodef{QoE}{Quality of Experience}
\acrodef{ACM}{Adaptive Coding and Modulation}
\acrodef{DAMA}{Demand Assignment Multiple Access}
\acrodef{VANET}{Vehicular Ad-Hoc Network}
\acrodef{RA}{Random Access}
\acrodef{CRA}{Contention Resolution ALOHA}
\acrodef{SA}{Slotted ALOHA}
\acrodef{DSA}{Diversity Slotted ALOHA}
\acrodef{OBU}{On-Board Unit}
\acrodef{CRDSA}{Contention Resolution Diversity Slotted ALOHA}
\acrodef{SIC}{Successive Interference Cancellation}
\acrodef{ARQ}{Automatic Repeat reQuest}
\acrodef{SC-ARQ}{Selective-Coded ARQ}
\acrodef{SR-ARQ}{Selective-Repeat ARQ}
\acrodef{IRSA}{Irregular Repetition Slotted ALOHA}
\acrodef{CGC}{Complementary Ground Component}
\acrodef{RSU}{Road Side Unit}
\acrodef{ACK}{Acknowledgment}
\acrodef{NACK}{Negative Acknowledgment}
\acrodef{DVB-SH}{Digital Video Broadcasting - Satellite Services to Handhelds}
\acrodef{DVB-H}{Digital Video Broadcasting - Handheld}
\acrodef{SACK}{Selective Acknowledgment}
\acrodef{SNACK}{Selective Negative Acknowledgment}\acrodef{SNACK}{Selective Negative Acknowledgment}
\acrodef{SNIR}{Signal to Noise plus Interference Ratio}
\acrodef{SCPS-TP}{Space Communications Protocol Specifications - Transport Protocol}
\acrodef{CCSDS}{Consultative Committee for Space Data Systems}
\acrodef{ESA}{European Space Agency}
\acrodef{NASA}{National Aeronautics and Space Administration}
\acrodef{BSM}{Broadband Satellite Multimedia}
\acrodef{RLNC}{Random Linear Network Coding}
\acrodef{NC}{Network Coding}
\acrodef{FIFO}{First In, First Out}
\acrodef{FCFS}{First Come, First Served}
\acrodef{BLER}{Block Error Rate}
\acrodef{GEO}{Geostationary Orbit}
\acrodef{LEO}{Low Earth Orbit}
\acrodef{FTP}{File Transfer Protocol}
\acrodef{CRC}{Cyclic Redundancy Check}
\acrodef{MAC}{Medium Access Control}
\acrodef{HTTP}{Hypertext Transfer Protocol}
\acrodef{ISP}{Internet Service Provider}
\acrodef{MSS}{Maximum Segment Size}
\acrodef{BIC}{Binary Increase Congestion control}
\acrodef{AQM}{Active Queue Management}
\acrodef{XCP}{eXplicit Control Protocol}
\acrodef{ECN}{Explicit Congestion Notification}
\acrodef{CA}{Congestion Avoidance}
\acrodef{RED}{Random Early Detection}
\acrodef{TD}{Triple-Duplicate}
\acrodef{TO}{Time Out}
\acrodef{IP}{Internet Protocol}
\acrodef{WMN}{Wireless Mesh Network}
\acrodef{ssthresh}{Slow-Start threshold}
\acrodef{MPE-IFEC}{Multi Protocol Encapsulation - Inter-burst Forward Error Correction}
\acrodef{FEC}{Forward Error Correction}
\acrodef{ML}{Maximum Likelihood}
\acrodef{CRC}{Cyclic Redundancy Check}
\acrodef{P2P}{Peer-to-Peer}
\acrodef{FMT}{Fade Mitigation Technique}
\acrodef{SGD}{Smart Gateway Diversity}
\acrodef{NCC}{Network Control Centre}
\acrodef{ModCod}{Modulation and Coding}
\acrodef{FIFO}{First-In-First-Out}
\acrodef{WRR}{Weighted Round Robin}
\acrodef{WFQ}{Weighted Fair Queuing}
\acrodef{NS}{Network Simulator}
\acrodef{GSE}{Generic Stream Encapsulation}
\acrodef{PDF}{Probability Density Function}
\acrodef{CDF}{Cumulative Density Function}
\acrodef{CoV}{Coefficient of Variation}
\acrodef{MSC}{Message Sequence Chart}
\acrodef{ESA}{European Space Agency}
\acrodef{LIU}{Lebanese International University}
\acrodef{TUM}{Technical University of Munich}
\acrodef{MSCE-CS}{Master of Science in Communications Engineering - Communications Systems}
\acrodef{DLR}{German Aerospace Center}
\acrodef{NC-SGD}{Network Coding for SGD}
\acrodef{ATSP}{Advanced Transport Satellite Protocol}
\acrodef{STP}{Satellite Transport Protocol}
\acrodef{WMN}{Wireless Mesh Network}
\acrodef{SNR}{Signal-to-Noise Ratio}
\acrodef{SINR}{Signal-to-Interference-plus-Noise Ratio}
\acrodef{LMS}{Land Mobile Satellite}
\acrodef{LTE}{Long-Term Evolution}
\acrodef{M2M}{Machine-to-Machine}
\acrodef{IoT}{Internet of Things}
\acrodef{RA}{Random Access}
\acrodef{UAV}{Unmanned Aerial Vehicle}
\acrodef{UAS}{Unmanned Aerial System}
\acrodef{FANET}{Flying Ad-Hoc Network}
\acrodef{MANET}{Mobile Ad-Hoc Network}
\acrodef{VANET}{Vehicle Ad-Hoc Network}
\acrodef{C2}{Command and Control}
\acrodef{DTN}{Delay Tolerant Network}
\acrodef{COTS}{Commercial Off-the-Shelf}
\acrodef{GCS}{Ground Control Station}
\acrodef{CSMA/CA}{Carrier Sense Multiple Access with Collision Avoidance}
\acrodef{SHE}{Smart Healthy Environment}
\acrodef{MSN}{Mobile Social Network}
\acrodef{WSN}{Wireless Sensor Network}
\acrodef{RSS}{Received Signal Strength}
\acrodef{GPS}{Global Positioning System}
\acrodef{GPRS}{General Packet Radio Service}
\acrodef{PAN}{Personal Area Network}
\acrodef{D2D}{Device to Device}
\acrodef{PMF}{Probability Mass Function}
\acrodef{ICT}{Information and Communications Technology}

\acrodefplural{ICT}[ICTs]{Information and Communications Technologies}
\acrodefplural{RSU}[RSUs]{Road Side Units}
\acrodefplural{VANET}[VANETs]{Vehicular Ad-Hoc Networks}
\acrodefplural{OBU}[OBUs]{On-Board Units}
\acrodefplural{UAV}[UAVs]{Unmanned Aerial Vehicles}
\acrodefplural{FANET}[FANETs]{Flying Ad-Hoc Networks}
\acrodefplural{MANET}[MANETs]{Mobile Ad-Hoc Networks}
\acrodefplural{VANET}[VANETs]{Vehicle Ad-Hoc Networks}
\acrodefplural{UAS}[UASs]{Unmanned Aerial Systems}
\acrodefplural{PMF}[PMFs]{Probability Mass Functions}

%% file: intro-FD.tex
\section{Introduction}
\label{sec:intro}
The idea of Smart Cities (SCs) has its roots in the so-called \textit{Healthy Cities} programme \cite{who}, launched in 1987 by the World Health Organization (WHO) and still active. It is a long-term international development initiative aimed at placing citizens\textquotesingle \space health in urban areas high on the agendas of the decision makers, to promote comprehensive local strategies for health protection and sustainable development. Currently, WHO is focusing on the so-called Health 2020 programme (phase VI), exploiting \textit{whole-of-government} and \textit{whole-of-society} approaches. The latter approach explicitly takes into account the views of citizens, by considering and acknowledging the relationship between people and governors, in order to deal with needs and issues as soon as they arise.
In the framework of SCs, the air quality is one of the fundamental factors that should be constantly monitored because of its effects on health \cite{chen2007ozone, chen2007PM, Brook2002} and, more generally, on citizens\textquotesingle \space Quality of Life (QoL). In fact, WHO launched the \textit{Global Platform on Air Quality and Health}, calling for a collaborative effort in order to develop, implement and monitor air pollution abatement strategies.
In this context, the \ac{SHE} project designed and deployed an ecosystem of Information and Communication Technology (ICT) solutions aimed at monitoring the environmental conditions and actively including citizens in the generation and exploitation of useful information. The first objective was to develop a cost-effective, distributed and efficient sensor network for collecting, processing and distributing data related to the air quality in the city of Pisa, Italy. It consists of both fixed and mobile sensor nodes able to measure several environmental parameters. Through the use of the mobile nodes, we investigated the feasibility of a totally mobile sensor network for urban environmental monitoring and the impact of the mobility of the nodes on the measurement accuracy. 
Mobile sensor nodes offer several advantages: (\textit{i}) they can extend the coverage of the environmental monitoring at a very low cost, thus complementing and/or substituting data derived from fixed nodes; (\textit{ii}) a high spatial and temporal resolution is achievable, providing a statistically reliable measurements flow; (\textit{iii}) the maintenance of the nodes is simplified, not requiring interventions of specialised technicians in  potentially distant locations; (\textit{iv}) the malfunctioning of a single sensor node can be simply identified  by comparing its measurements with those provided by close devices. In order to evaluate whether or not the mobility of nodes can affect the quality of the measurements, a real test-bed has been conducted in the city of Pisa by comparing the measurements collected by fixed and mobile sensor nodes built within the \ac{SHE} project.

As a second goal, the sensor network extends the environmental monitoring by exploiting the personal mobile devices of the citizens, allowing them to report and comment on those situations that can further influence the environmental conditions and, more in general, QoL in the city. Specifically, a \ac{MSN} application, namely \textit{SmartCitizen}, has been developed to implement the participatory sensing paradigm, by directly collecting and sharing contents provided by citizens through their personal mobile devices (i.e., smartphones and tablets). Citizens provide subjective information (e.g., comments, pictures, videos) that represents an additional support to the objective information collected by the sensor nodes. 
In fact, user-generated contents are both distributed among citizens through the opportunistic communication of their mobile devices, and stored on the central host of the herein proposed architecture, in order to be correlated with data collected by nodes. Then, this information is made available to citizens, city governors, local authorities, and all the other experts interested in receiving timely updates on the air quality in the city. 

In order to make the information accessible to citizens in a simple way, we defined three indexes derived from sensor data: Air Quality Index (AQI), Thermal Comfort Index (TCI), and Traffic Index (TI). The last one is obtained by an additional sensor network already present in the city of Pisa and managed by a local company, \textit{Pisa Mobility} 
(PisaMO). 
Accessing the project website\footnote{\ac{SHE} website is available at \url{http://www.progettoshe.it}}, indexes are represented as simple colored indicators. In addition, experts can directly access raw data, in order to carry detailed analyses and to support the decision process; experts can be decision makers (for instance, local governors) and researchers, which can take advantage of data in order to identify long-period trends and to support scientific analyses. 
The \ac{SHE} project also included a medical study on the impact of environmental pollution on citizens' health. Anyway, this study is still in progress and, for this reason, it is not described in this work.

The rest of this paper is organized as follows: Section \ref{sec:rw} provides a survey of the literature in the context of SCs, with a focus on pollution monitoring systems.
Section \ref{sec:sc} describes the reference scenario and the proposed system architecture, along with the sensor nodes in use. Then, it provides details on the air quality, thermal comfort and traffic indexes and the developed mobile app. The experimental results are provided in Section \ref{sec:res} and, finally, lessons learned and conclusions are in Section \ref{sec:conclusions}.

%% file: related-FD.tex
\section{Background and rationale of this work}
\label{sec:rw}
In the last few years several solutions for environmental monitoring has been presented in literature.
The authors in \cite{baralis2016analyzing} propose an air pollution analysis based on meteorological and traffic data collected in Milan during 2013. The following parameters were collected: (\textit{i}) pollutant concentrations, including Particulate Matter (PM) 10 and 2.5, carbon monoxide (CO), ozone (O$_3$), nitrogen dioxide (NO$_2$), benzene (C$_6$H$_6$), and sulfur dioxide (SO$_2$); (\textit{ii}) climate conditions, such as air temperature, relative humidity, precipitation level, wind speed and atmospheric pressure; (\textit{iii}) traffic data for different categories of vehicles. Monitoring stations were deployed along the city; meteorological data were gathered from a simple personal weather station and, finally, traffic data were collected thanks to the use of \acp{WSN}. A web-based interface was developed to provide access to the data and the results of the data analyses. However, the whole system in \cite{baralis2016analyzing} is based on the use of fixed installations only.
The use of mobile sensor nodes has been recently studied in \cite{xiang2016poster}, where sensors are placed on rentable bicycles. Every time a bicycle is in use, the sensor collects data and stores them on the available local storage; when the bicycle is returned, the collected data, geo-referenced thanks to a \ac{GPS} receiver, are sent to a central server via a \ac{GPRS} connection.  The authors aimed at demonstrating that the measurements collected by the mobile sensor nodes do not depend on the sensor orientation and on the sampling period; instead, the actual path is crucial to correctly evaluate the results, especially to estimate the distribution of pollutants in a target area. However, the authors could not effectively demonstrate the accuracy of the mobile measurements as the samples collected by the mobile sensor nodes were compared to indoor measurements, hardly comparable with outdoor ones.
Crowd-sourced air quality monitoring is studied in \cite{yang2016towards}, where also the physical activity of volunteers has been recorded, in order to assist environmental health studies. Low-cost mobile sensor nodes have been used during a campaign that lasted half a year in the city of Gj{\o}vik, Norway, proving the feasibility of a low-cost crowd-sourced data collection platform. Anyway, the authors only proved the feasibility of such a platform, since the campaign had a limited duration.
In \cite{brienza2015low}, the authors propose an air quality monitoring system, namely \textit{uSense}, based on the use of small and low-cost nomadic devices, running on batteries. Those sensor nodes can be moved from one place to another when the need for readings in a new area arises: only O$_3$, NO$_2$, and CO are monitored and transferred via a wireless connection to a central server. The server estimates an air quality index, accessible to users via a Web interface or a mobile application.

The main difference between our solution and those previously cited consists in the realization of a complete ecosystem for environmental monitoring, relying on a distributed \ac{WSN}, composed of both fixed and mobile nodes, and on the participatory sensing paradigm, by directly involving citizens through their personal mobile devices, according to a \textit{whole-of-society} approach. The ecosystem could also be extended with wearable devices aimed at further increasing the coverage area of the monitoring system and also at providing a different perspective in the collection of environmental data, e.g., to analyse the individual and collective exposure to pollutants. A first attempt to use these type of body-worn sensors has been proposed in \cite{nikzad2012citisense} by using a prototype sensor board. Currently, some emerging companies are investing in the development of such devices trying to guarantee a comparable level of accuracy with respect to certified sensor stations. However, they are still not available in commerce at this moment and, for this reason, we decided to initially focus on the explicit user contribution to enrich sensor data as fundamental part of participatory sensing paradigm. In addition, SHE ecosystem is open to the integration of heterogeneous devices and external monitoring systems implementing the OGC SWE standard\footnote{Open Geospatial Consortium - Sensor Web Enablement standard http://www.opengeospatial.org/ogc/markets-technologies/swe} for sensor data encoding. To this aim, SmartCitizen app supports the same standard directly on mobile devices by integrating an apposite software framework to optimise sensor data management with respect to the limited resources of mobile devices \cite{arnaboldi2013sensor}. \ac{SHE} data is also accessible in customised ways, according to the users' categories (i.e., common citizens, experts, local authorities), both through the mobile app and the project website, in order to ensure a global comprehension of the system's results.

%% file: scenario-FD.tex
\section{Scenario and System Architecture}
\label{sec:sc}
\begin{figure*}[!h]
	\begin{center}
		\includegraphics[scale=0.55, clip=true, trim=0 0 0 0]{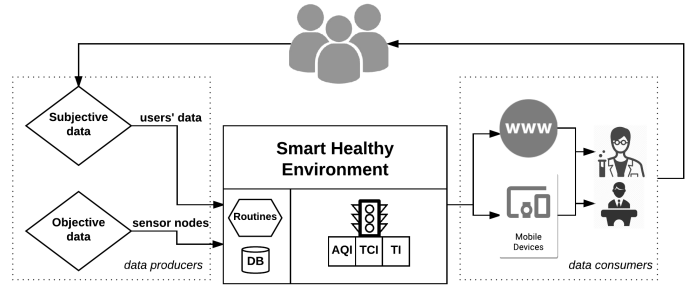}
	\end{center}
	\caption{A logical description of the \ac{SHE} architecture: objective and subjective data are collected from fixed and mobile sensor nodes, in order to derive air quality indicators, presented to different types of users via website and mobile application. The citizens are the core of the system, representing both data producers and consumers.}
   	\label{fig:she}
\end{figure*}

The system architecture we designed is depicted in Figure \ref{fig:she}: objective (derived from sensors) and subjective (generated by citizens) data are collected from heterogeneous nodes in order to derive useful indicators and to generate a space for discussion among citizens and between citizens and local governors. By only relying on the objective data, AQI, TCI and TI indexes are computed and subsequently presented to different user categories: citizens, scientists, and local authorities.

\begin{table}[ht]
\centering
\begin{tabular}{cl}
\multicolumn{1}{l}{\includegraphics[scale=0.2, clip=true, trim=0 0 0 0]{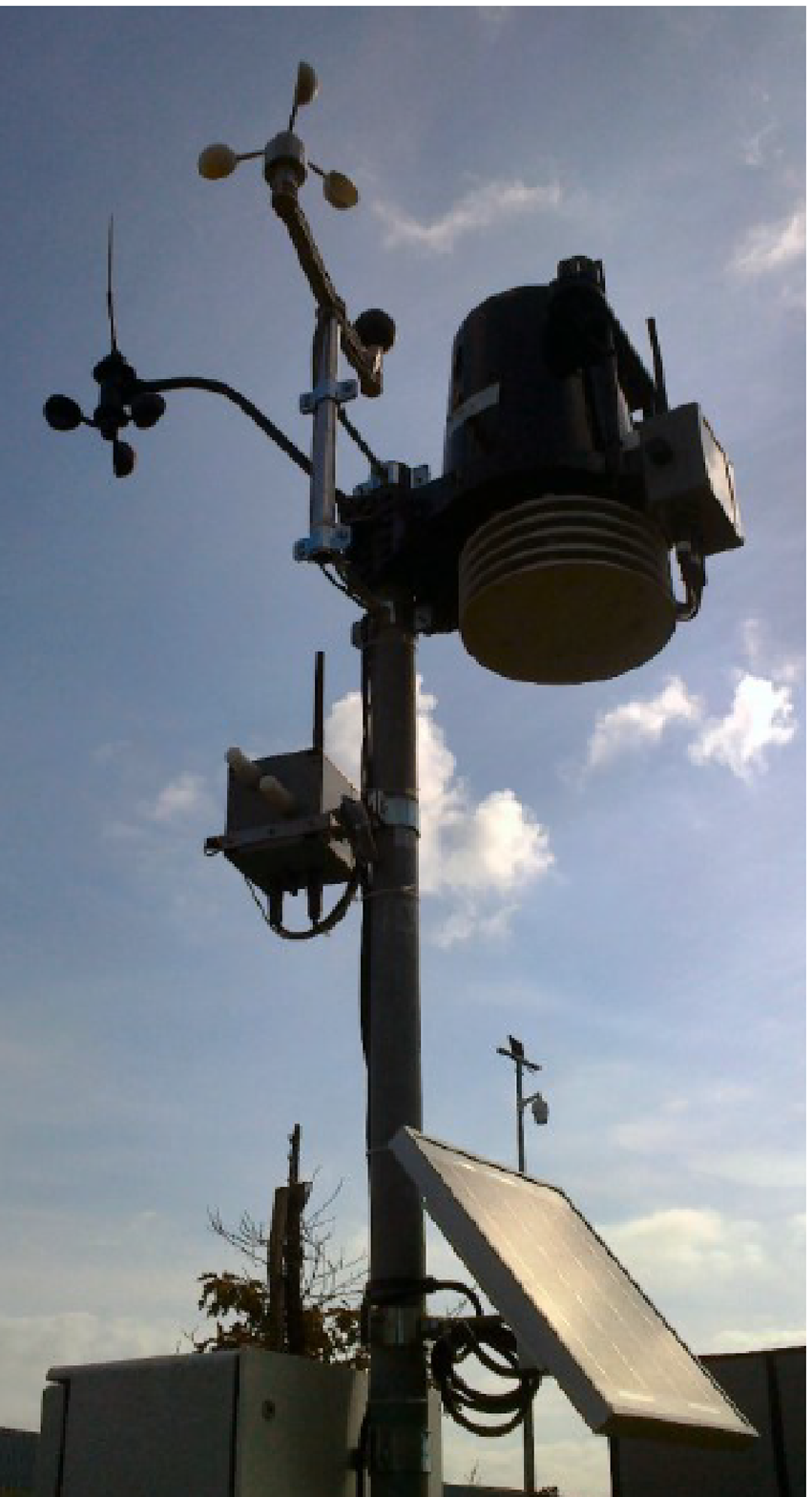}} & \includegraphics[scale=0.37, clip=true, trim=0 450 450 0]{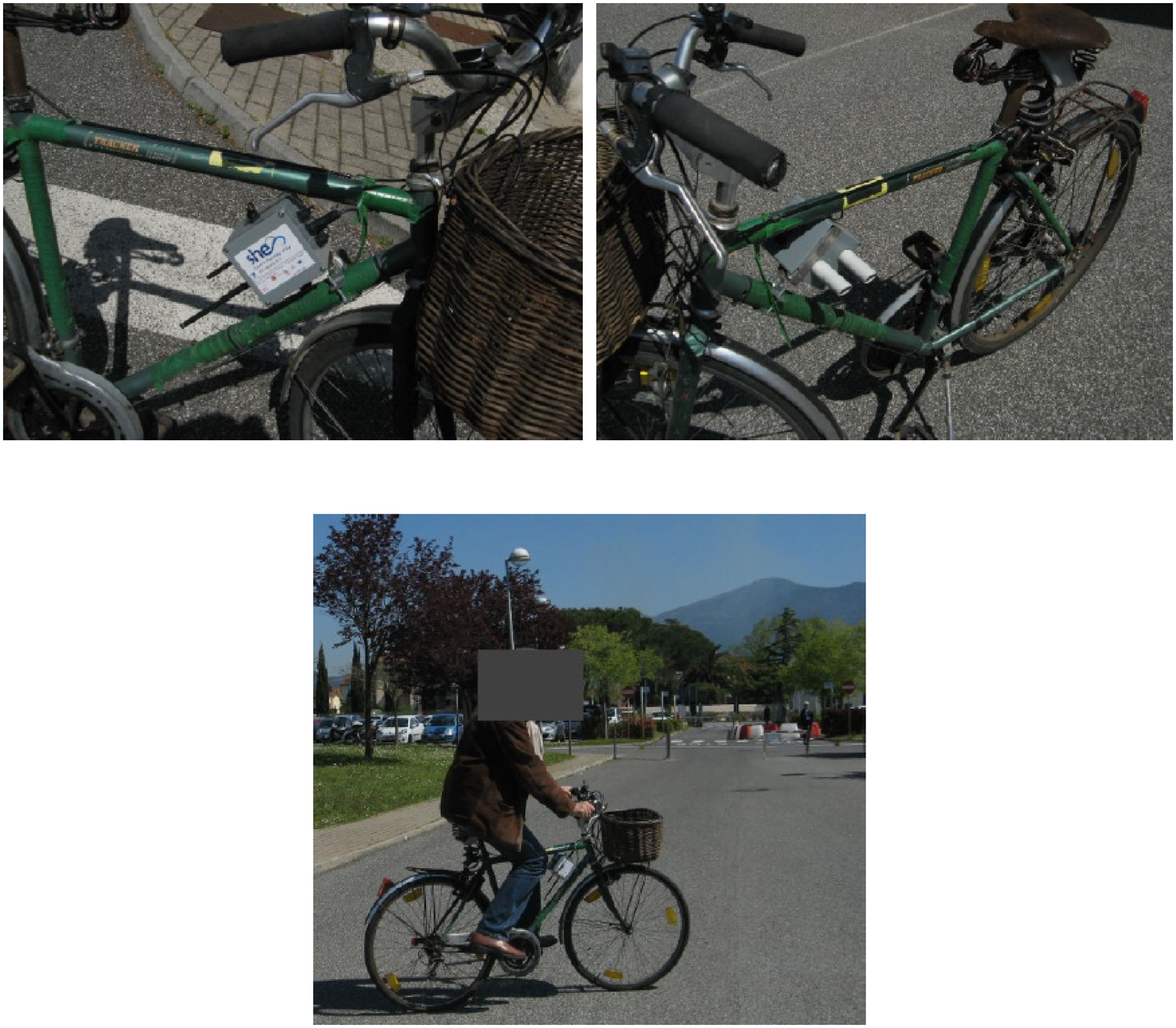} \\
\begin{tabular}[c]{@{}c@{}}(a) The fixed\\weather station\end{tabular} & \multicolumn{1}{c}{\begin{tabular}[c]{@{}c@{}}(b) Sensor node installed\\on a bicycle\end{tabular}} \\
\multicolumn{2}{c}{\begin{tabular}[c]{@{}c@{}}Fig. 2. The sensor node and the weather station in use in\\ the proposed architecture for monitoring the air quality.\end{tabular}}
\end{tabular}
\end{table}

\begin{figure*}
	\centering
    \includegraphics[scale=0.3, clip=true, trim=0 0 0 0]{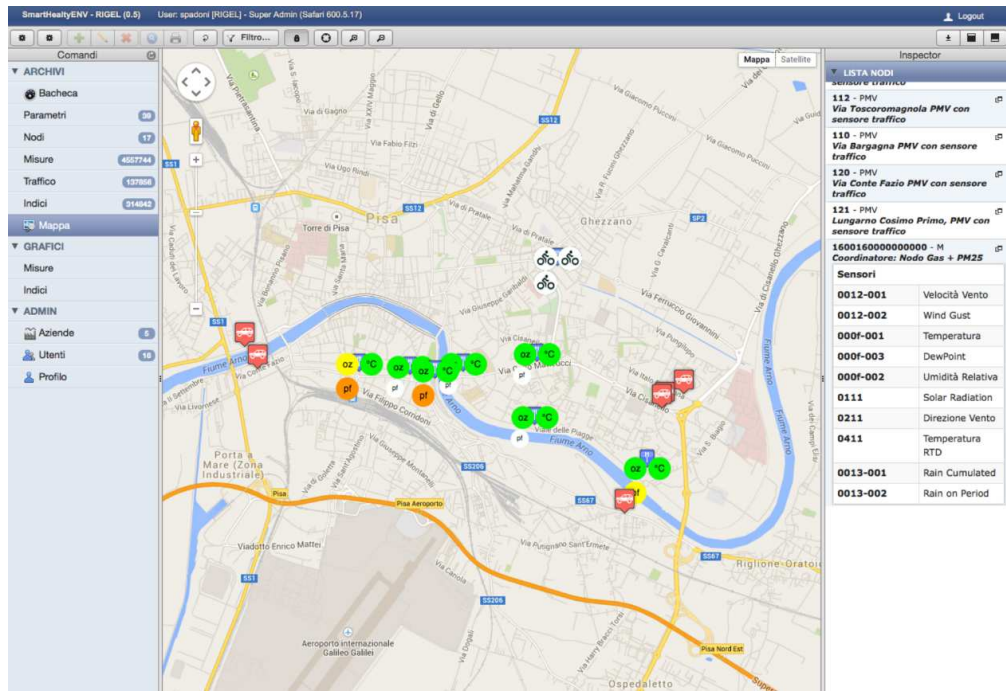}
	\caption{A screenshot of the \ac{SHE} website: AQI and TCI indexes are plotted as colored circles; the mobile nodes are shown\\as a white icon surrounding a cycle. The traffic sensors are squared, red filled and surrounding a car.}
	\label{fig:website}
\end{figure*}

The objective data are then complemented with the subjective data generated by citizens through the SmartCitizen \ac{MSN} application.
Our sensor platform collects the following environmental parameters: CO, carbon dioxide (CO$_2$), unburned hydrocarbons (HC), O$_3$ and PM 2.5 pollutants, in addition to micro-meteorological parameters. $n=9$ sensor nodes have been deployed in the city of Pisa: seven are fixed ones, two are mobile ones; furthermore, we also used a weather station along with a solar panel to recharge internal batteries (see Figure 2a).
The sensor node we developed has been used as fixed and mobile one, and is visible in Figure 2b, installed on a bicycle.
The fixed sensor nodes have been placed on two intersecting paths in the city center: (\textit{i}) a \textit{heavy traffic} path, thus expecting higher pollutants concentrations, (\textit{ii}) and a \textit{fitness} path, a low-traffic path close to the city center and used by citizens for outdoor fitness activity.
A sensor node used as network coordinator has been placed at the intersection point of the two paths. The distance between two consecutive fixed nodes is approximately 350 meters on the heavy traffic path and approximately one kilometer on the fitness path, so that the whole network covers an area of approximately 4 to 5 square kilometers, corresponding to Pisa\textquotesingle s city center.
Figure \ref{fig:website} shows the position of the fixed, mobile and traffic sensor nodes, as visible on the website. Several colored icons are shown per physical node (refer to Section \ref{sec:sensors}).
The traffic sensors are placed at the entrance of the main arterial roads from and towards the city center. The instantaneous position of the mobile sensor nodes is visible in the top center of Figure \ref{fig:website}: they are identified by a white icon surrounding a cycle.
The base sensor node is adaptable and easily expandable: it is provided with large processing capacity, a Wireless M-Bus (WMBus) \cite{hersentm} radio in the case of fixed nodes, WMBus and ZigBee \cite{baronti2007wireless} radios in the case of mobile nodes, a solar cell to recharge the internal battery, and programmable analogue and digital components to interface with the on-board sensors.
WMBus has been used for the data exchange among the fixed nodes. ZigBee has joined the WMBus protocol in the case of mobile nodes, aiming at studying the operative limits of the ZigBee protocol when used in vehicular contexts at urban speeds. The mobile nodes have been installed on bicycles traveling around the city (but cars and buses can be used, too), as for instance visible in Figure 2b. 
We also performed an additional study by using an \ac{UAV} carrying a simplified sensor node as payload, in order to study the quality of the transmission channel. The results of the latter study are reported in \cite{bacco2014uavs, bacco2014radio}. 
The communication infrastructure allows both the communication among the nodes and the communication between the coordinator and the central host (acting as server), where the collected data are stored and processed. The central server is hosted at a remote location, and it provides a database service for storing collected data, analysis procedures, control interfaces for experts and technicians, and a public open interface accessible via the aforementioned website.

\subsection{Sensors and communication platform}
\label{sec:sensors}
We identified the following sensors for the detection of the aforementioned environmental parameters:
\begin{itemize}
\item Sensirion SHT75 (temperature, relative humidity and dew point);
\item Non-Dispersive Infrared Sensor (NDIR) gas sensor (CO, CO$_2$, HC), namely G3, based on Dual Wavelength Ratioing (DWR) technology;
\item SGX Sensortech MiCS-2614 (O$_3$);
\item Davis Instruments 7911 and WS1070/WS (anemometers);
\item Davis Instruments 7859 (rain gauge);
\item Arcus RPTF 2 PT1000 (radiant temperature);
\item Qbit MP25 (PM 2.5);
\item Davis Instruments 6490 (radiometer);
\item Freescale MPL3115A2 (atmospheric pressure);
\item \ac{GPS} SoC Mediatek MT3329 (mobile nodes only).
\end{itemize}
The specification of the aforementioned commercial sensors (apart from the G3 one) can be found online. G3 sensor has been specifically designed and built within the project activities, because no commercial product was found, at that time, satisfying the system requirements. Electrochemical sensors have been excluded because of their limited average lifespan, in favor of NDIR technology.
The G3 sensor performs four IR measurements in four different optic bands, and the response time (T90) is of the order of tens of seconds ($< 1.5$ minutes). In order to have accurate measurements, the gas sensor must reach the optimal exercise temperature, and it may take up to 15 minutes before the first reading is available. The limit of detections (LoDs) are: $<5$ppm for CO, $<10$ppm for CO$_2$, and $<5$ppm for HC, with a resolution of 1ppm for CO, 1ppm for CO$_2$, and 1ppm for HC. The node is powered with a lithium battery operating at 3.7 V. Qbit Optronics developed the gas sensor according to the provided specifications, while the control electronics of the gas sensor have been designed and realized by project partners. 
All sensor nodes have been calibrated in laboratory before use and have been re-calibrated every three months, on average. The G3 sensor node built within the project has been calibrated (one point calibration at $\approx$70\% of the span value) by relying on calibration gas cylinders.

To summarize, three types of nodes have been used:  
\begin{itemize}
\item a weather station node: measuring temperature, relative humidity, dew point, wind speed, directions and gusts, heavy rain, solar radiation;
\item seven fixed nodes: measuring temperature, relative humidity, dew point, wind speed and gusts, PM 2.5\footnote{Only three of the seven fixed nodes measure PM 2.5.}, HC, CO$_2$, CO, O$_3$, gas temperature, gas relative humidity, gas atmospheric pressure, radiant temperature.
\item two mobile nodes: measuring temperature, relative humidity, dew point, HC, CO$_2$, CO, O$_3$, gas temperature, gas relative humidity, gas atmospheric pressure.
\end{itemize}
Each fixed or mobile node measures the aforementioned pollutants every $T_N = 5$ minutes and immediately transmits the results to the coordinator node, which transmits the collected measurements every $T_I = 15$ minutes to the central server via a \ac{GPRS} connection. Therefore, $n T_I / T_N$ reports are received by the central server every $T_I$ minutes, triggering the update of the aforementioned indexes. The fixed nodes rely on the WMBus protocol (operating at 169Mhz) to exchange data with the coordinator, while the mobile nodes rely on the WMBUS/ZigBee protocol (the latter operating at 2.4 GHz) if in the communication range of another sensor node or, alternatively, on a \ac{GPRS} connection.

\subsection{Air Quality, Thermal Comfort and Traffic indexes}
\label{subsec:indexes}
As already mentioned, we defined a set of indexes aimed at providing some clear indications on the city's environmental conditions. 
\subsubsection{AQI}
information on the air quality is conveyed to the citizens thanks to two sub-indexes\footnote{We adopted WHO guidelines available at: \url{http://www.who.int/mediacentre/factsheets/fs313/en/}.}: the first one, namely AQI$_{O3}$, is based on the ozone measurements; the second one, namely AQI$_{PM}$, is based on PM 2.5 measurements. AQI$_{O3}$, according to WHO guidelines, should be less than a mean concentration of 100 $\mu$g$/m^3$ within an 8-hour time window; AQI$_{PM}$, according to WHO guidelines, should be less than a yearly mean concentration of 10 $\mu$g$/m^3$.
The sub-indexes are updated by applying a moving average every 8 hours in the case of O$_3$, and every 24 hours in the case of PM 2.5. 
Table \ref{tab:aqi} provides details on the thresholds in use for evaluating AQI sub-indexes, as shown in Fig. \ref{fig:website} (AQI$_{O3}$ is marked as \textit{oz} and AQI$_{PM}$ is marked as \textit{pf} in Fig. \ref{fig:website}). The value of the indexes does not provide any forecast. 
\begin{table}
\centering
   \begin{tabular}{|c|c|} \hline
   \textbf{Thresholds} & \textbf{Index color}\\ \hline \hline
      O$_3 <$ 100					& green		\\ \hline
      100 $\leq$ O$_3 <$ 180 		& yellow 	\\ \hline
      180 $\leq$ O$_3 <$ 240 		& orange	\\ \hline
      O$_3$ $\geq$ 240 				& red 		\\ \hline \hline

	  PM 2.5 $<$ 10	 				& green	 	\\ \hline
      10 $\leq$ PM 2.5 $<$ 25	 	& yellow 	\\ \hline
      25 $\leq$ PM 2.5 $<$ 60	 	& orange 	\\ \hline
      PM 2.5 $\geq$ 60 				& red 		\\ \hline
	\end{tabular}
	\caption{Thresholds for AQI$_{O3}$ and AQI$_{PM}$ sub-indexes.\\Both O$_3$ and PM 2.5 readings are measured in $\mu$g$/m^3$.}
   \label{tab:aqi}
\end{table}

\subsubsection{TCI}
it is based on the Universal Thermal Comfort Index (UTCI), which is a readily accessible thermal index based on a state-of-the-art thermo-physiological model \cite{jendritzky2012utci}. Several applications and services are based on the use of such an index, for instance Public Weather Services and Public Health Systems, as well as precautionary planning and climate impact research in the health sector \cite{jendritzky2012utci}. TCI value is updated every 15 minutes, using as an input the air temperature, the mean radiant temperature, the wind speed, and the relative humidity. The index is presented in the same color-coded way of AQI index; in Figure \ref{fig:website}, it is marked as \textit{\SI{}{\degreeCelsius}}, and its thresholds can be read in Table \ref{tab:utci}. Also in this case, the value of the index does not provide any forecast.
\begin{table}
\centering
   \begin{tabular}{|c|c|} \hline
   \textbf{Temperature ranges} & \textbf{Index color}\\ \hline \hline
   	  $38 \leq T < 46$ 		& dark red			\\ \hline
      $32 \leq T < 38$ 		& red				\\ \hline
      $26 \leq T < 32$ 		& orange			\\ \hline
      $9 \leq T < 26$ 		& green 			\\ \hline
      $0 \leq T < 9$ 		& blue				\\ \hline
      $-13 \leq T < 0$ 		& dark blue 		\\ \hline \hline
	\end{tabular}
	\caption{Thresholds for TCI index.\\The temperature $T$ is measured in \textit{\SI{}{\degreeCelsius}}.}
   \label{tab:utci}
\end{table}

\subsubsection{TI}
the index in (\ref{eq:traffic}) estimates the traffic in cities assuming that a continuous flow of vehicles passes a \textit{virtual line}, in the following referred to as \textit{access}, where the traffic is measured. The index has the following formulation:
\begin{equation}
	TI = s_b \thinspace K_1 \thinspace K_2 \thinspace K_3 \thinspace K_4 \thinspace \thinspace \thinspace [EV / s]
    \label{eq:traffic}
\end{equation}
where $s_b \approx 1800$ is a constant value, representing a base congestion factor, while $K_i$ factors are used to adjust the latter to the actual situation. The index is measured in Equivalent Vehicles (EV) per second. The most common vehicle classes (or types) are reported in Table \ref{tab:classes}, according to the Italian regulations. More specifically, $K_1 = \dfrac{1}{\sum_i{a_i \thinspace E_i}}, \sum_i{a_i} = 1$ describes the composition of the traffic (cars, motorcycles, trucks), as reported in Table \ref{tab:classes}; $K_2 = 1 \pm 0.03 \thinspace i$ takes into account the steepness $i \%$ of the access: it decreases in case of uphill and increases in case of downhill; $K_3$ considers the specific position of the access in the urban area (for instance, centre or suburbs) and Table \ref{tab:localization} reports its possible values; $K_4 = \dfrac{1}{\sum_i{b_i \thinspace G_i}}, \sum_i{b_i} = 1$ considers the interference due to the presence of pedestrians and to the maneuvering of vehicles, as reported in Table \ref{tab:manev}.
Twenty-two points for measuring traffic flows are placed around the city of Pisa, tracking vehicles entering and leaving the city, as well as the class of the vehicles.
\begin{table*}
\centering
\begin{tabular}{ccc}
  \begin{subtable}[t]{.2\textwidth}
	\centering
   \begin{tabular}{|c|c|} \hline
   \textbf{Vehicle classes} & \textbf{E}\\ \hline \hline
      bicycles & 0.2 \\ \hline
      motorcycles & 0.33 \\ \hline
      cars & 1 \\ \hline
      trucks & 1.75 \\ \hline
      buses & 2.25 \\ \hline
      trams & 2.5 \\ \hline
  \end{tabular}
  \caption{Vehicle classes}
  \label{tab:classes}
  \end{subtable}
  &
  \begin{subtable}[t]{.2\textwidth}
  \centering
   \begin{tabular}{|c|c|} \hline
   \textbf{Localization} & \textbf{$K_3$}\\ \hline \hline
    residential & 1 \\ \hline
    commercial & 0.98 \\ \hline
	industrial & 0.93 \\ \hline
	business & 0.85 \\ \hline
   \end{tabular}
   \caption{Localization parameter} 
   \label{tab:localization}
   \end{subtable}
   &
   \begin{subtable}[t]{.2\textwidth}
   \centering
   \begin{tabular}{|c|c|} \hline
   \textbf{Maneuvering type} & \textbf{G}\\ \hline \hline
    straight line & 1 \\ \hline
    turning right & 1 - 1.25 \\ \hline
	turning left & 1 - 1.75 \\ \hline
    \end{tabular}
    \caption{Maneuvering types}
    \label{tab:manev}
\end{subtable}
\end{tabular}
\caption{Values of the parameters used for the estimation of the Traffic Index (TI)}
\end{table*}

\subsection{SmartCitizen \ac{MSN} application}
\label{subsec:msn}
\begin{figure*}
	\centering
    \begin{subfigure}[t]{0.65\textwidth}
	\includegraphics[scale=0.23, clip=true, trim=0 0 0 0]{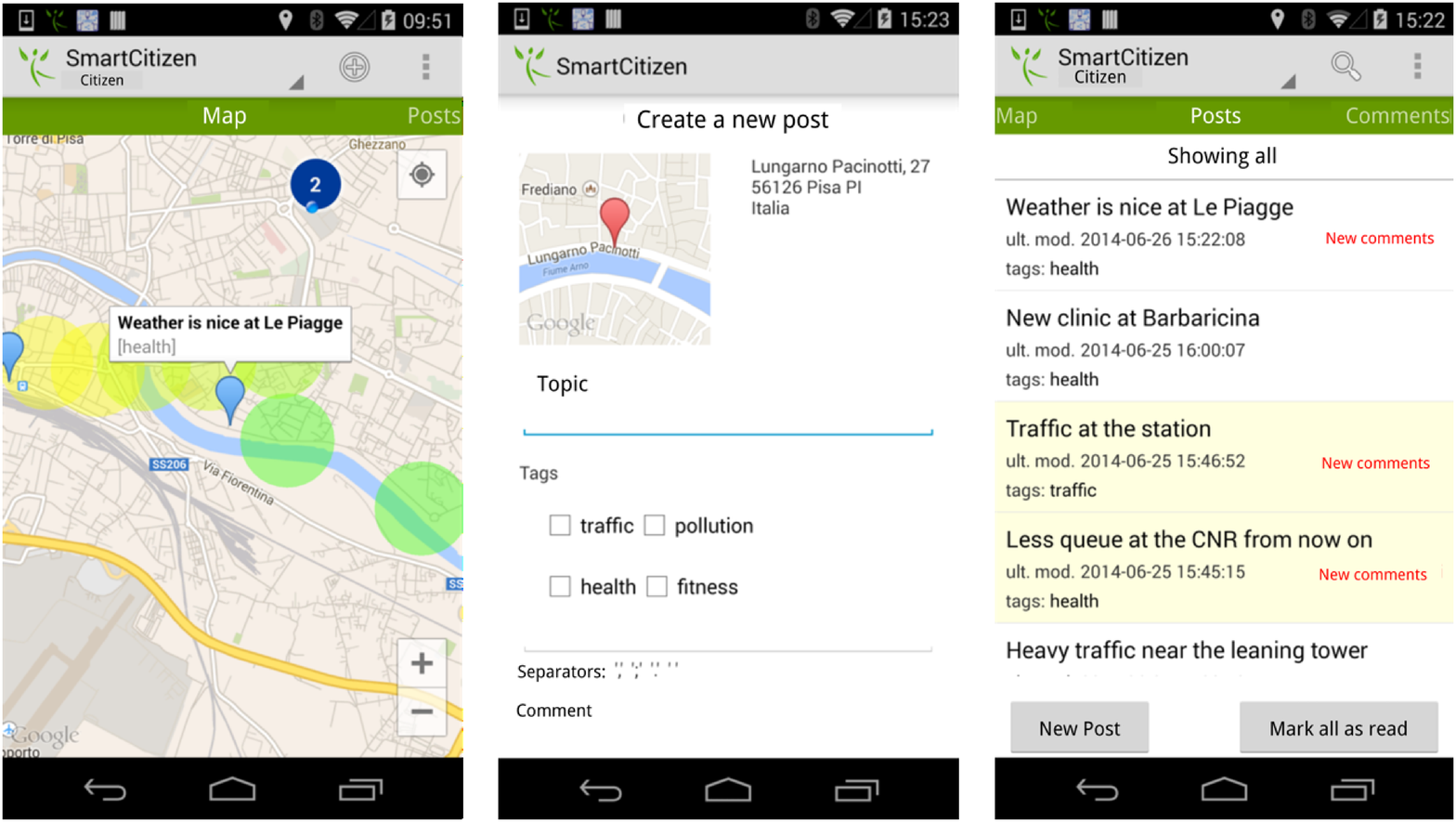}
    \end{subfigure}
    \begin{subfigure}[t]{0.32\textwidth}
	\includegraphics[scale=0.3, clip=true, trim=0 0 0 0]{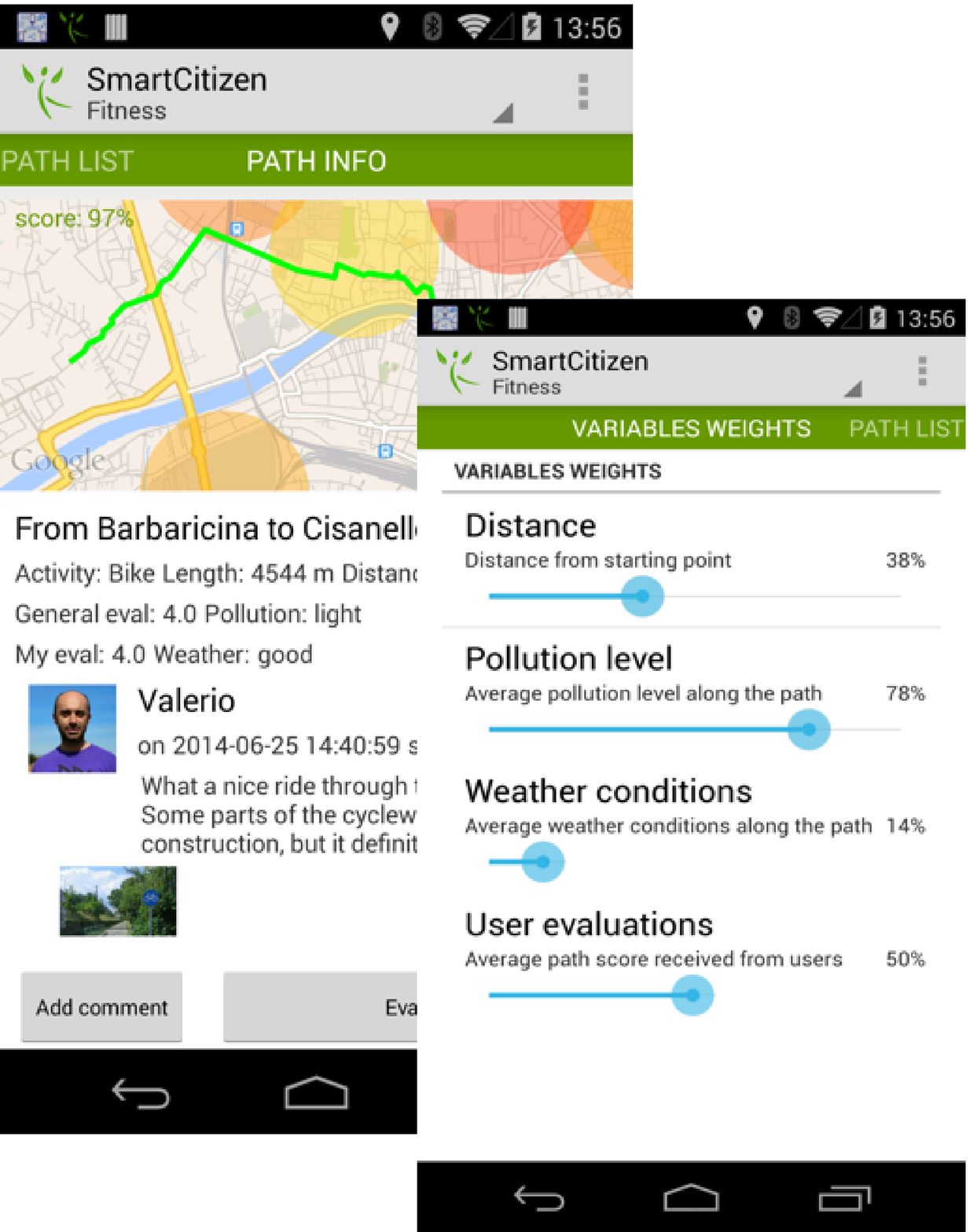}
    \end{subfigure}
    \caption{Some screenshots of the \ac{MSN} application, namely \textit{SmartCitizen}, developed inside the \ac{SHE} project}
    \label{fig:msn}
\end{figure*}
The SmartCitizen app aims at stimulating the active participation of citizens in collecting and sharing useful data related to QoL in their city, not only on the environmental conditions, but also on concurrent events or experiences that can provide additional information on the environmental situation. The app intuitively visualizes data provided by the deployed \ac{WSN}, which provides the indexes discussed in Section \ref{subsec:indexes}, in a general and simple presentation consisting of colored circles, centered on the current geographical position of the sensor nodes. The app also stimulates discussions among citizens on several topics exploiting features similar to social networking applications, such as posting, commenting and chatting. The main difference between SmartCitizen and standard Online Social Networks is that citizens share their contents and experiences via \ac{D2D} communications, relying on proximity and opportunistic communications (as  the main principles of \ac{MSN} applications \cite{delmastro2016people}) and avoiding a continuous storage on a centralised infrastructure. Figure \ref{fig:msn} shows some screenshots of the graphical user interface of the app. In the first screenshot, the app presents the air quality index on a map, as measured on each sensing station deployed in the city, and presented as colored circles. The app is also designed for expert users able to interpret the detailed information about the measured quantities. In this case, the authorized users can visualize the detailed data of a station simply by clicking on the circle. All sensing data are downloaded from the central server, if the user device has Internet connectivity, or they can be downloaded through \ac{D2D} communication, if available on other users' devices in close proximity. The application also provides the user with the possibility of creating new posts, in which he can open discussion on a specific topic, tagging it for an optimised dissemination among users and devices.
Finally, the user can visualise the list of active discussions on different topics, and the list of generated contents for each discussion with appropriate notification of new available contents. We also integrated additional features designed for sports users, considering the increasing trend on outdoor fitness activities, the lack of useful information about running paths conditions and, overall, on the healthy conditions of specific city areas. 
Specifically, SmartCitizen allows users in close proximity to share fitness experiences in a quasi real-time scenario. In this case, they can directly record activity paths in the city, and enrich the data with their own information, comments, and suggestions. The air quality information in the area is automatically disseminated among the users' devices through a context-aware content dissemination protocol provided by the CAMEO middleware platform \cite{arnaboldi2014cameo}. SmartCitizen app also allows users to upload their contents to the aforementioned central server. This process is not fully automated, in order to limit the number of connections towards the server, thus forcing the user to upload \textit{significant} contents only. 
In terms of privacy, users can decide what piece of information has to be shared only locally (in close proximity) and what publicly (remotely stored). They must explicitly agree to a specific privacy statement and terms of use presented at the first application launch. Personal and sensible data are never exchanged. 

%% file: results.tex
\section{Experimental Results}
\label{sec:res}
The deployment of the described \ac{WSN} allowed us to conduct an extensive experimental evaluation of the feasibility of using mobile sensor nodes (instead of fixed stations) for environmental monitoring in Pisa, in addition to the evaluation of the air quality in the city. 
Mobile nodes are carried by a mobile entity (humans included), thus this type of network can provide accurate measurements on the individual exposure to single pollutants, further to environmental information on city areas. In the following, we focus on the comparison of data provided by (\textit{i}) fixed nodes on the heavy traffic and the fitness paths, and by (\textit{ii}) mobile and fixed nodes. The results of the statistical comparison are based on a measurement campaign in spring 2015 in Pisa. 
It is worth recalling that mobile and fixed nodes share the same hardware and software components for measuring the aforementioned pollutants, thus the overall setup is tested against mobility conditions.

\begin{figure}
    \centering
        \begin{subfigure}[t]{0.22\textwidth}
          \includegraphics[scale=0.36, clip=true, trim=0 0 0 0]{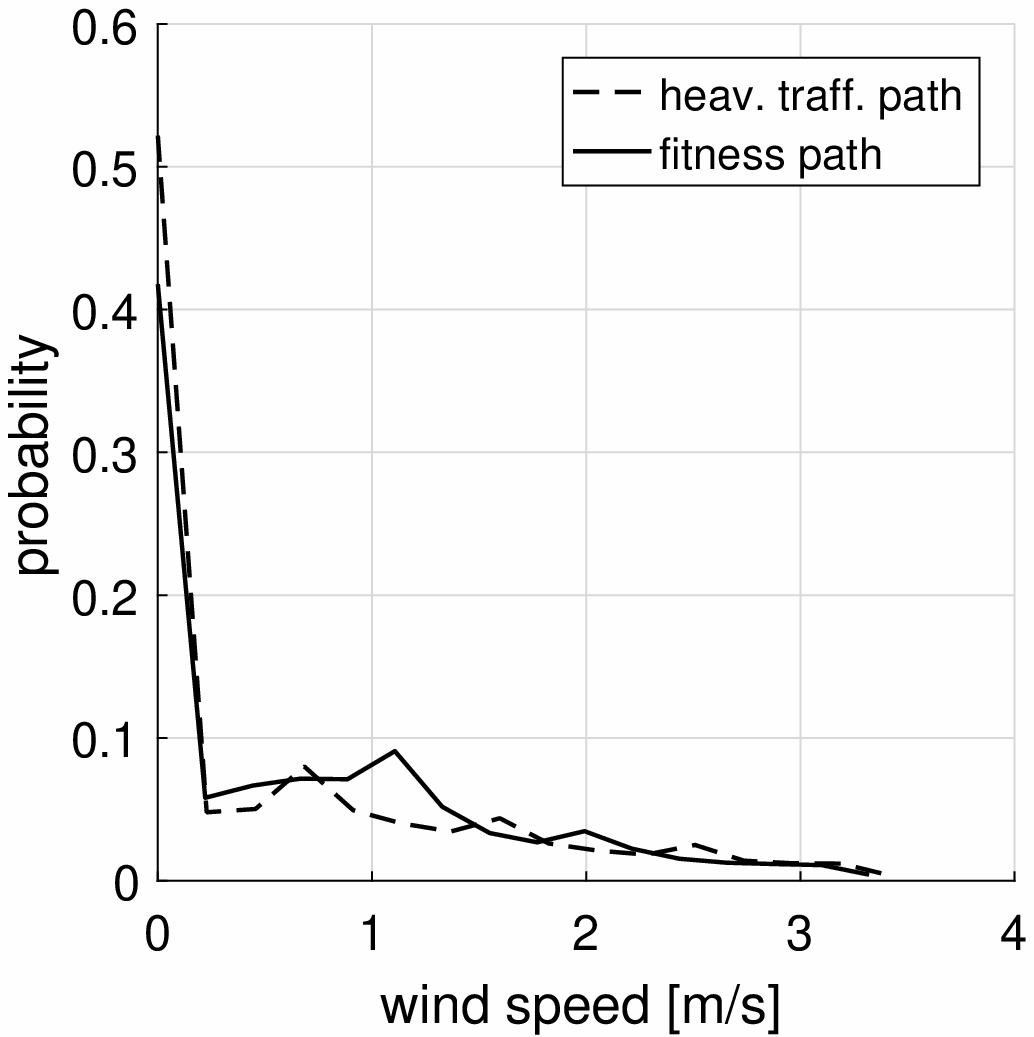}
          \label{fig:windspeedRG}
    	\end{subfigure}
    ~
        \begin{subfigure}[t]{0.22\textwidth}
          \includegraphics[scale=0.36, clip=true, trim=0 0 0 0]{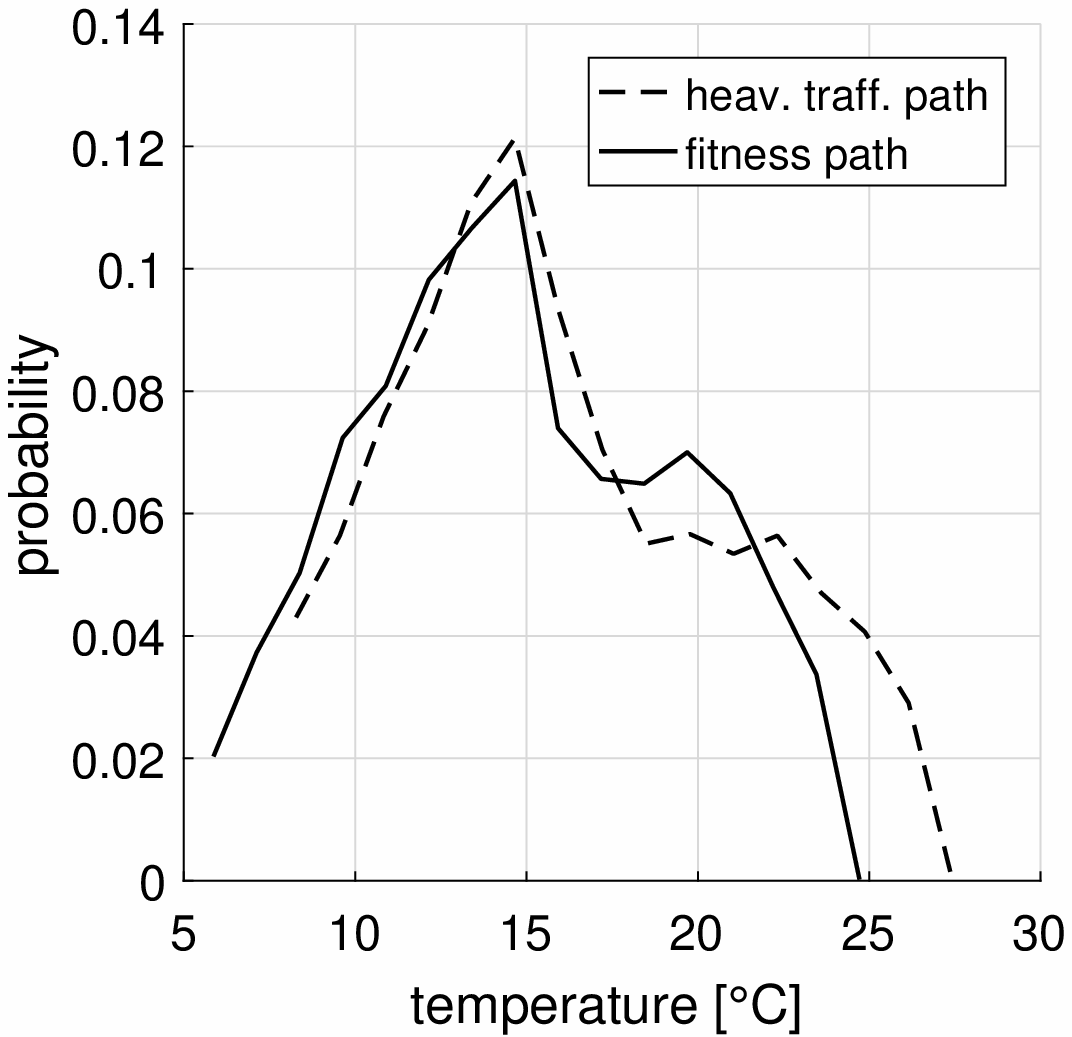}
          \label{fig:temperatureRG}
    	\end{subfigure}
    ~ 
        \begin{subfigure}[t]{0.22\textwidth}
          \includegraphics[scale=0.36, clip=true, trim=0 0 0 0]{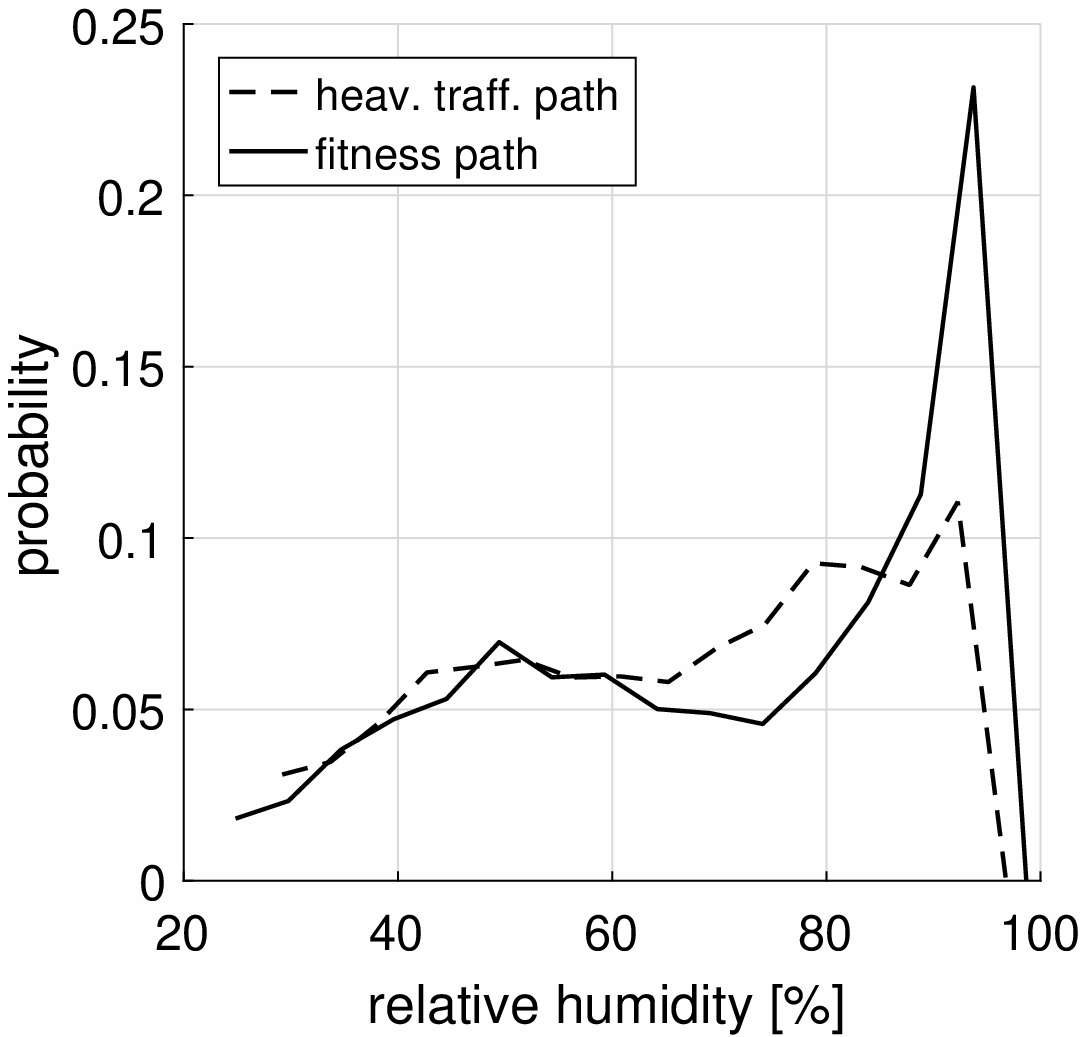}
          \label{fig:relhumidityRG}
    	\end{subfigure}
    ~ 
        \begin{subfigure}[t]{0.22\textwidth}
          \includegraphics[scale=0.36, clip=true, trim=0 0 0 0]{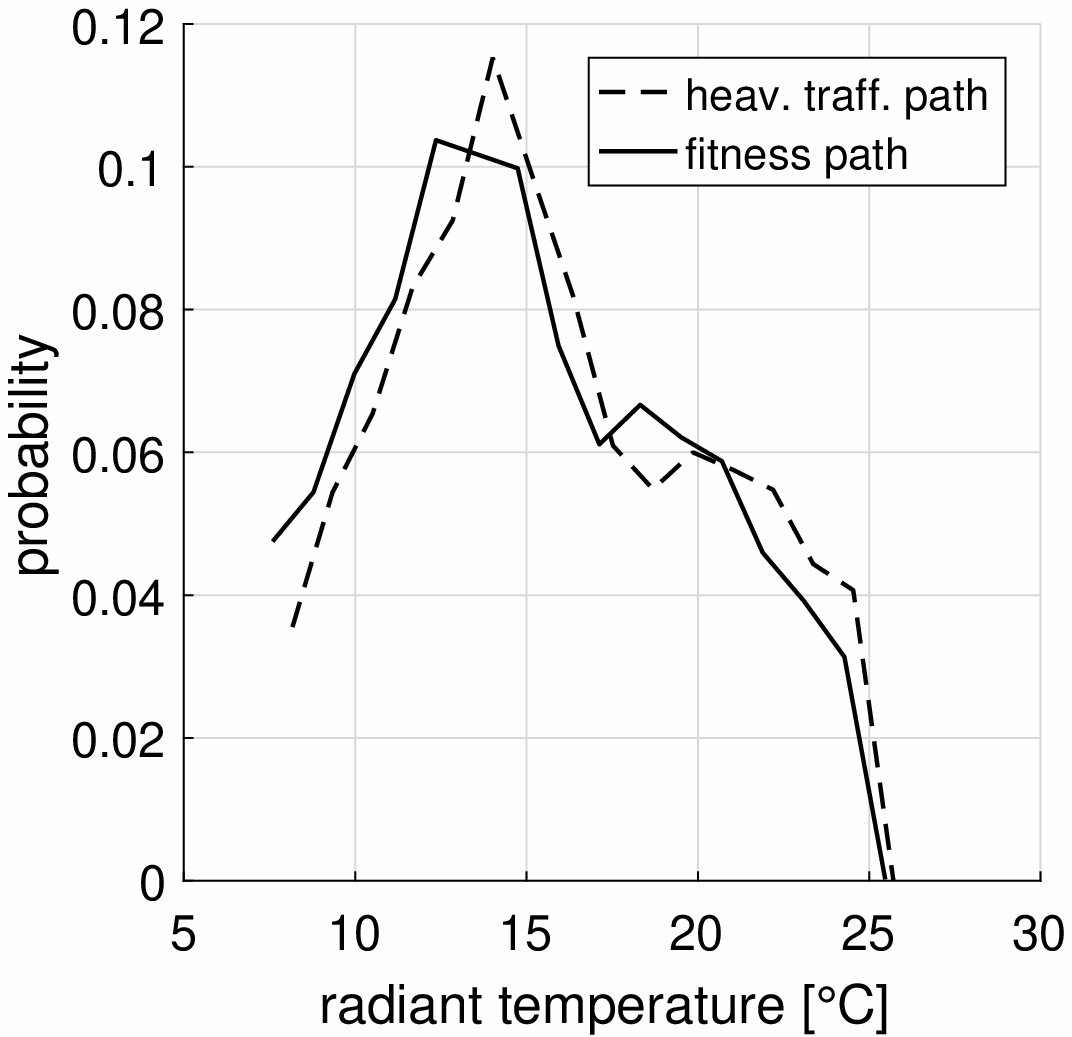}
          \label{fig:radiantRG}
    	\end{subfigure}
    ~
        \begin{subfigure}[t]{0.22\textwidth}
          \includegraphics[scale=0.36, clip=true, trim=0 0 0 0]{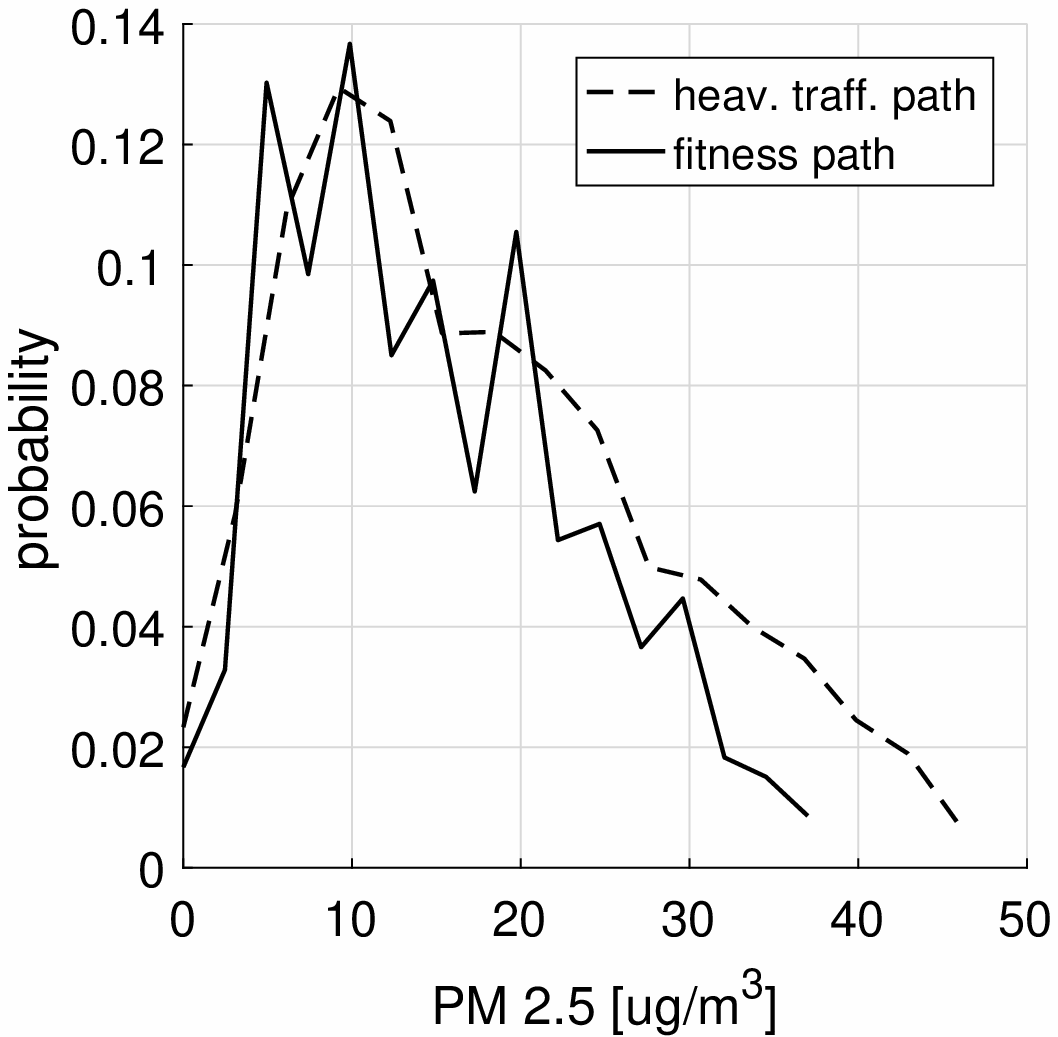}
          \label{fig:pmRG}
    	\end{subfigure}
        ~ 
        \begin{subfigure}[t]{0.22\textwidth}
          \includegraphics[scale=0.36, clip=true, trim=0 0 0 0]{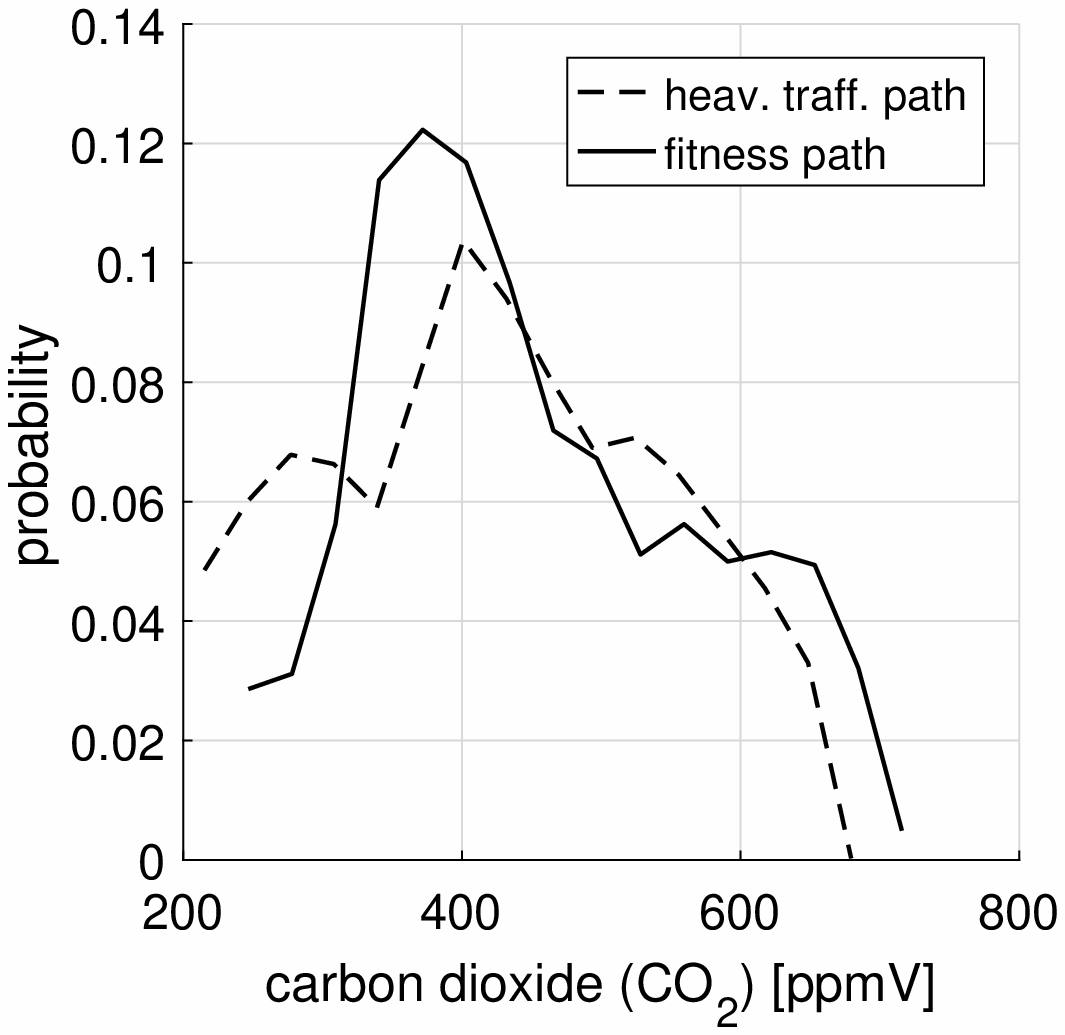}
          \label{fig:co2RG}
    	\end{subfigure}
        ~ 
        \begin{subfigure}[t]{0.22\textwidth}
          \includegraphics[scale=0.36, clip=true, trim=0 0 0 0]{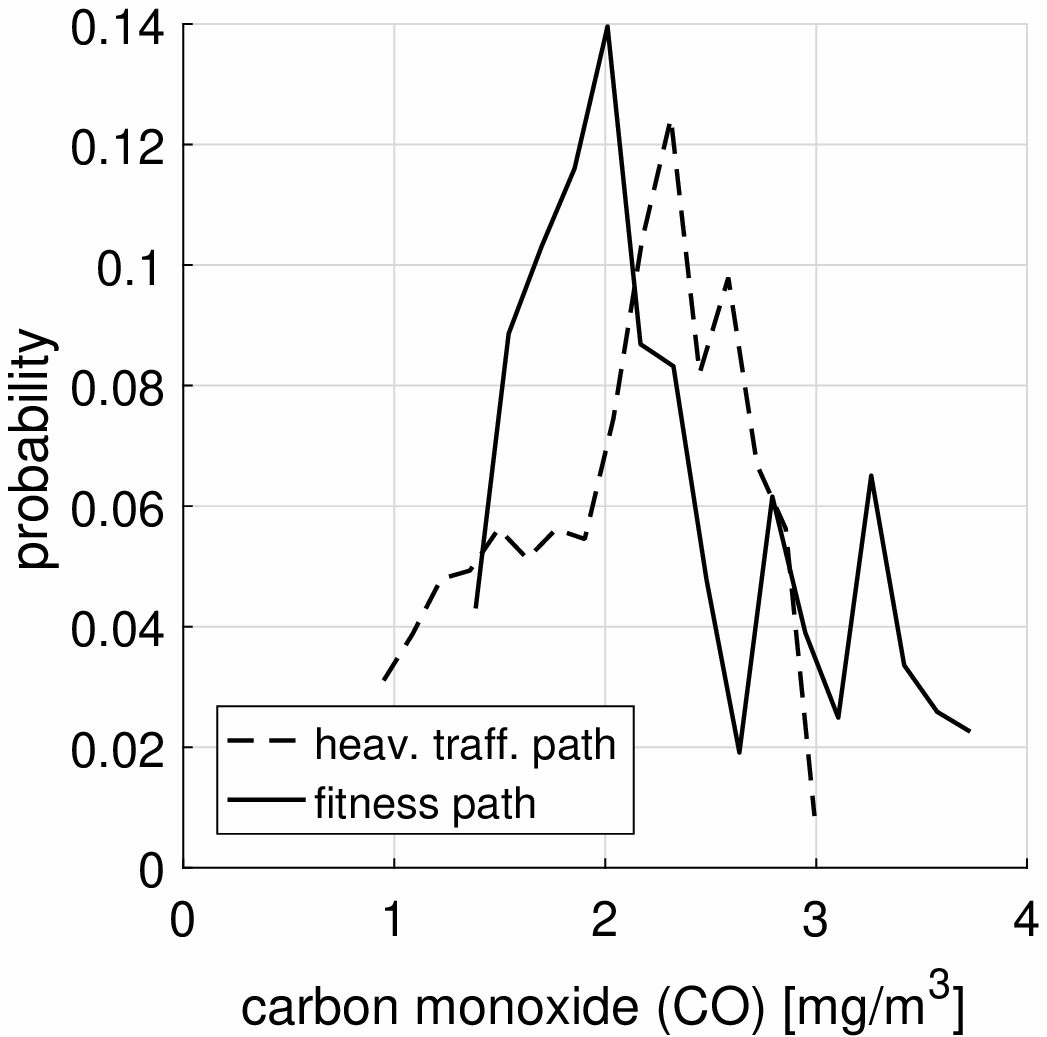}
          \label{fig:coRG}
    	\end{subfigure}
        ~ 
        \begin{subfigure}[t]{0.22\textwidth}
          \includegraphics[scale=0.36, clip=true, trim=0 0 0 0]{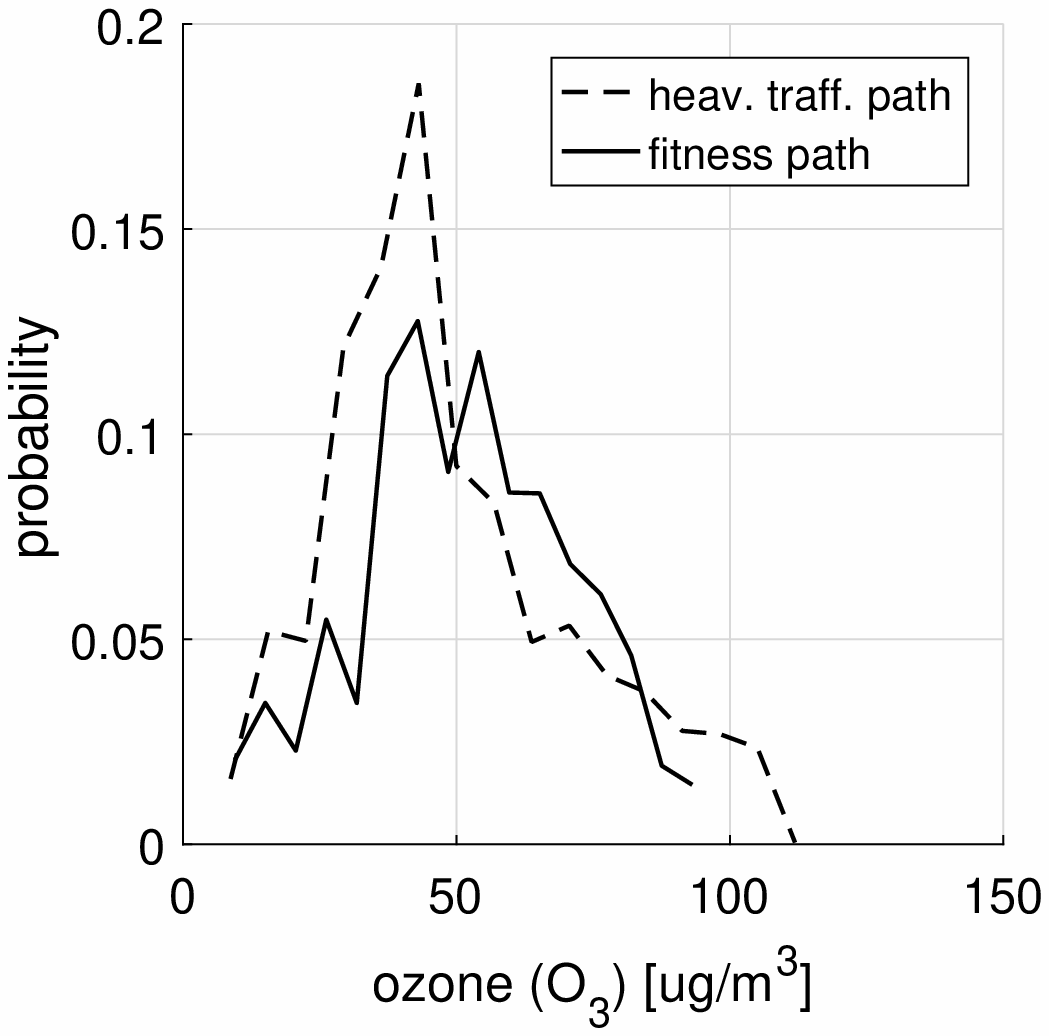}
          \label{fig:ozoneRG}
    	\end{subfigure}        
    \caption{Comparison among \acp{PMF} of the measurements on the heavy traffic and on the fitness paths}
    \label{fig:comparisonRG}
\end{figure}

Figure \ref{fig:comparisonRG} shows the comparison between \acp{PMF} of the considered pollutants and quantities collected by the sensor nodes located on the heavy traffic path and on the fitness path. The average values of \acp{PMF} in Figure \ref{fig:comparisonRG} are reported in Table \ref{tab:comparisonRG}, which also compares the readings by relying on the relative error definition: \mbox{$\eta_{TF} = \left| 1-\dfrac{M_{tr}}{M_{fit}} \right|$}. The \acp{PMF} of the dew point and of HC are not visible in Figure \ref{fig:comparisonRG} for space reasons ($\eta_{TF}$ is negligible for both quantities). 
Heavy traffic and fitness paths show similar average values because Pisa's centre is quite small.
\begin{table}
\begin{center}
  \resizebox{.99\hsize}{!}{
  \begin{tabular}{|c|c|c|c|} \hline
   \textbf{Measured quantity $M$} & \textbf{$M_{tr}$} & \textbf{$M_{fit}$} & \textbf{$\eta_{TF}$} \\ \hline \hline
      wind speed & 0.62 m/s & 0.69 m/s & 0.12 \\ \hline
      temperature & \SI{16.2}{\degreeCelsius} & \SI{14.7}{\degreeCelsius} & 0.09 \\ \hline
      relative humidity & 66.4\% & 70.2\% & 0.057 \\ \hline
      dew point & \SI{9.7}{\degreeCelsius} & \SI{9.8}{\degreeCelsius} & 0.01 \\ \hline
      radiant temperature & \SI{15.8}{\degreeCelsius} & \SI{15.1}{\degreeCelsius} & 0.04 \\ \hline
      PM 2.5 & 17.7 $\mu$g$/m^3$ & 14.8 $\mu$g$/m^3$ & 0.16 \\ \hline
      unburned hydrocarbons (HC) & 3.12 ppmV & 3.12 ppmV & 0.0006 \\ \hline 
      carbon dioxide (CO$_2$) & 423.26 ppmV & 451.1 ppmV & 0.06 \\ \hline
      carbon monoxide (CO) & 2.05 $m$g$/m^3$ & 2.28 $m$g$/m^3$ & 0.11 \\ \hline 
      ozone (O$_3$) & 48.97 $\mu$g$/m^3$ & 51.33 $\mu$g$/m^3$ & 0.05 \\ \hline 
  \end{tabular}
  }
\end{center}
\caption{Comparison between the average values collected on the heavy traffic ($M_{tr}$) and on the fitness ($M_{fit}$) paths.}
\label{tab:comparisonRG}
\end{table}

\begin{figure}
    \centering
        \begin{subfigure}[t]{0.22\textwidth}
          \includegraphics[scale=0.36, clip=true, trim=0 0 0 0]{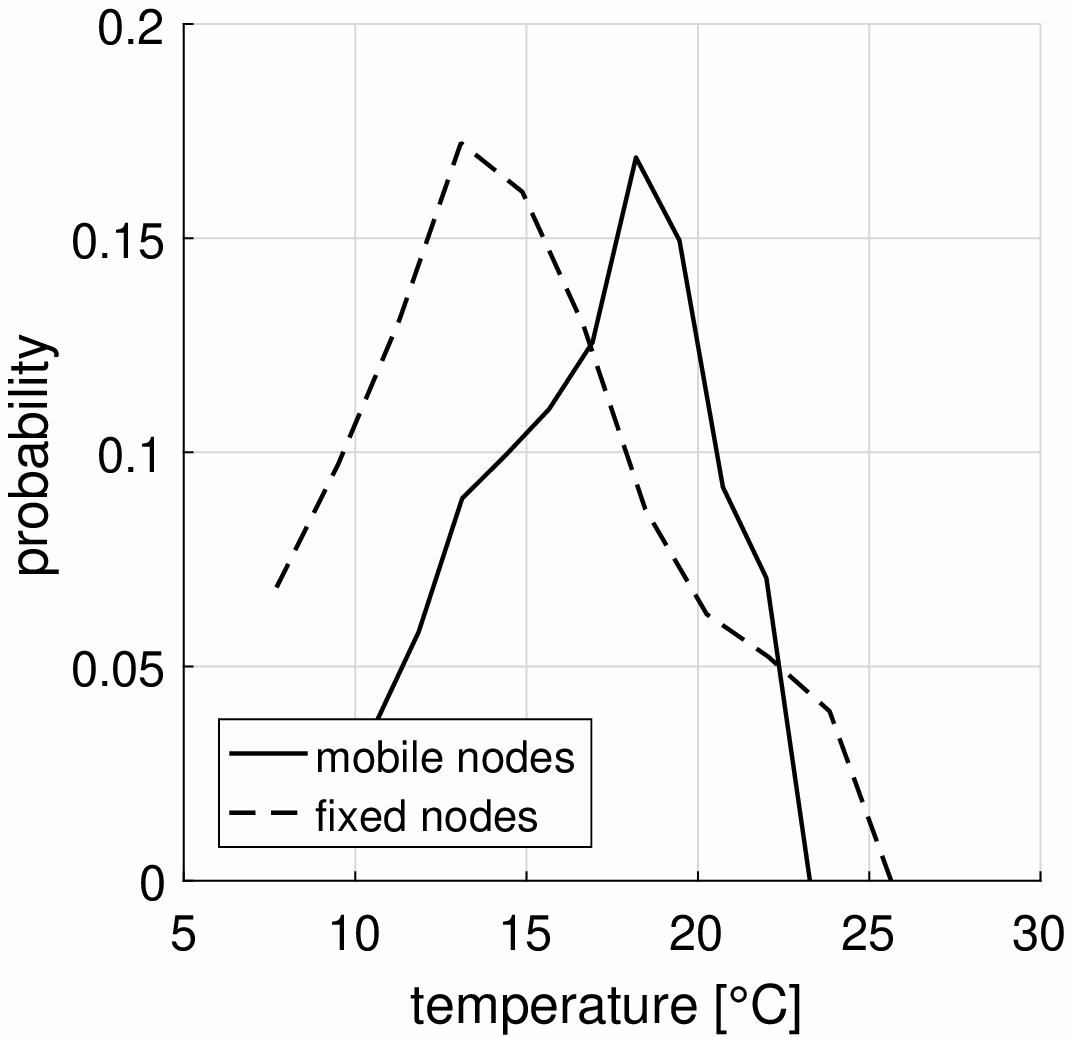}
          \label{fig:temperature}
    	\end{subfigure}
    ~
        \begin{subfigure}[t]{0.22\textwidth}
          \includegraphics[scale=0.36, clip=true, trim=0 0 0 0]{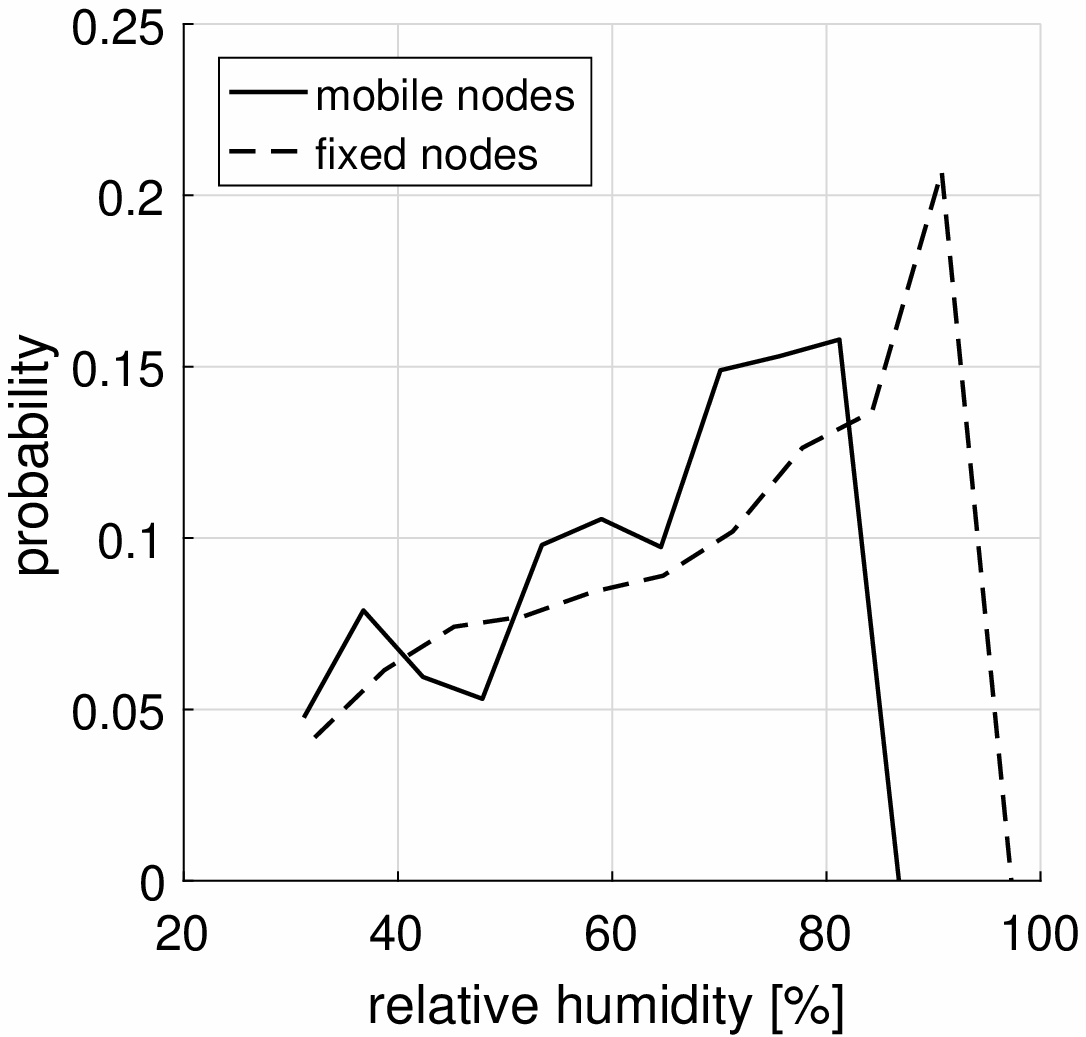}
          \label{fig:relative_humidity}
    	\end{subfigure}
    ~ 
        \begin{subfigure}[t]{0.22\textwidth}
          \includegraphics[scale=0.36, clip=true, trim=0 0 0 0]{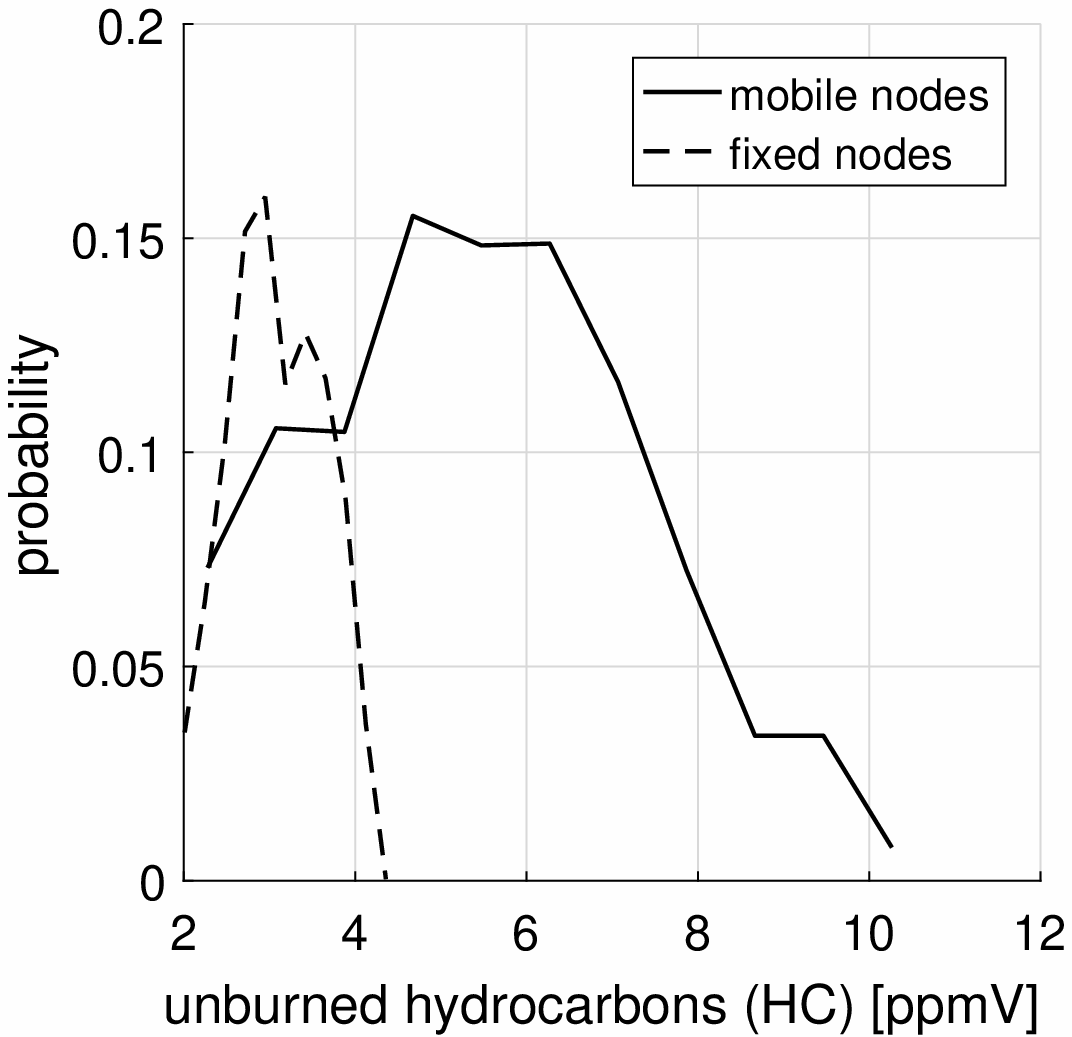}
          \label{fig:unburned_hydrocarbons}
    	\end{subfigure}
    ~ 
        \begin{subfigure}[t]{0.22\textwidth}
          \includegraphics[scale=0.36, clip=true, trim=0 0 0 0]{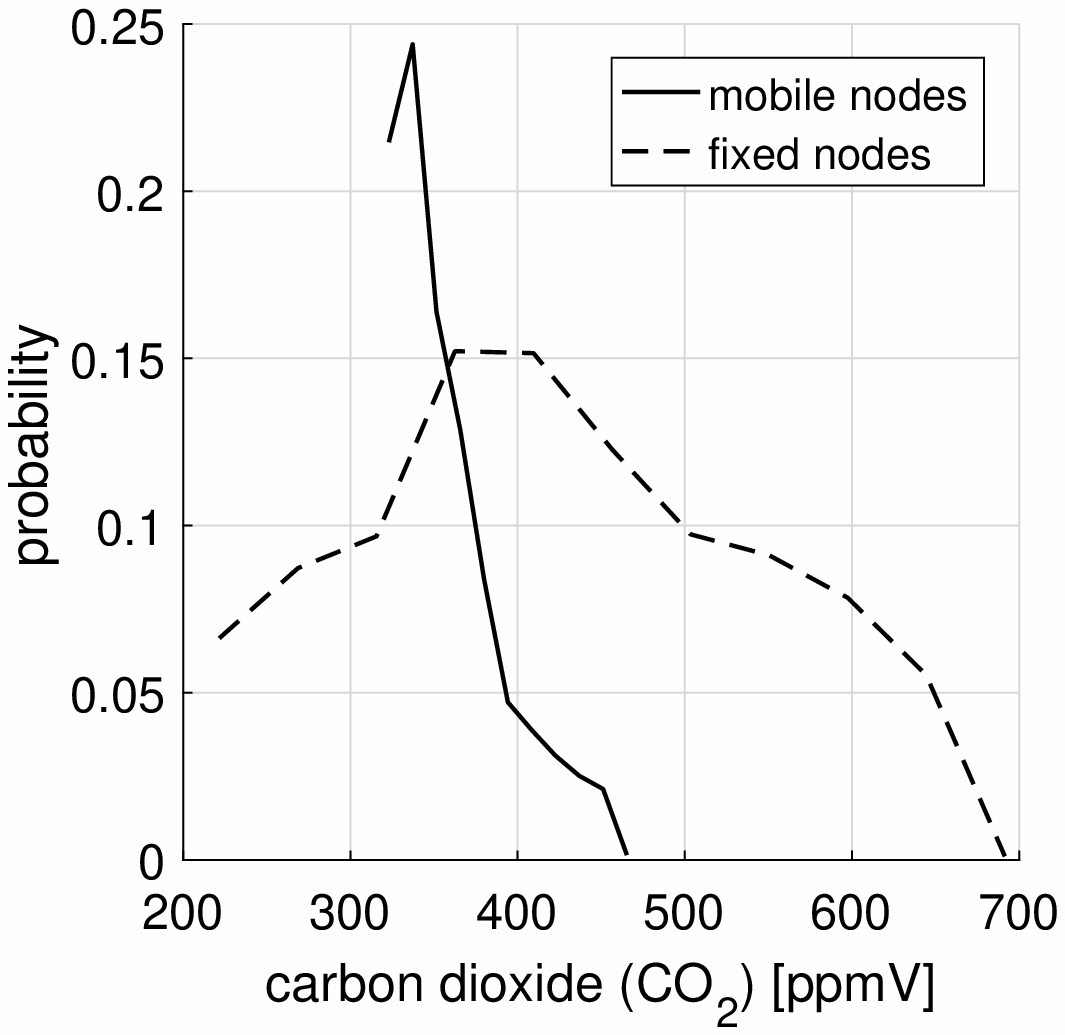}
          \label{fig:carbon_dioxide}
    	\end{subfigure}
    ~
        \begin{subfigure}[t]{0.22\textwidth}
          \includegraphics[scale=0.36, clip=true, trim=0 0 0 0]{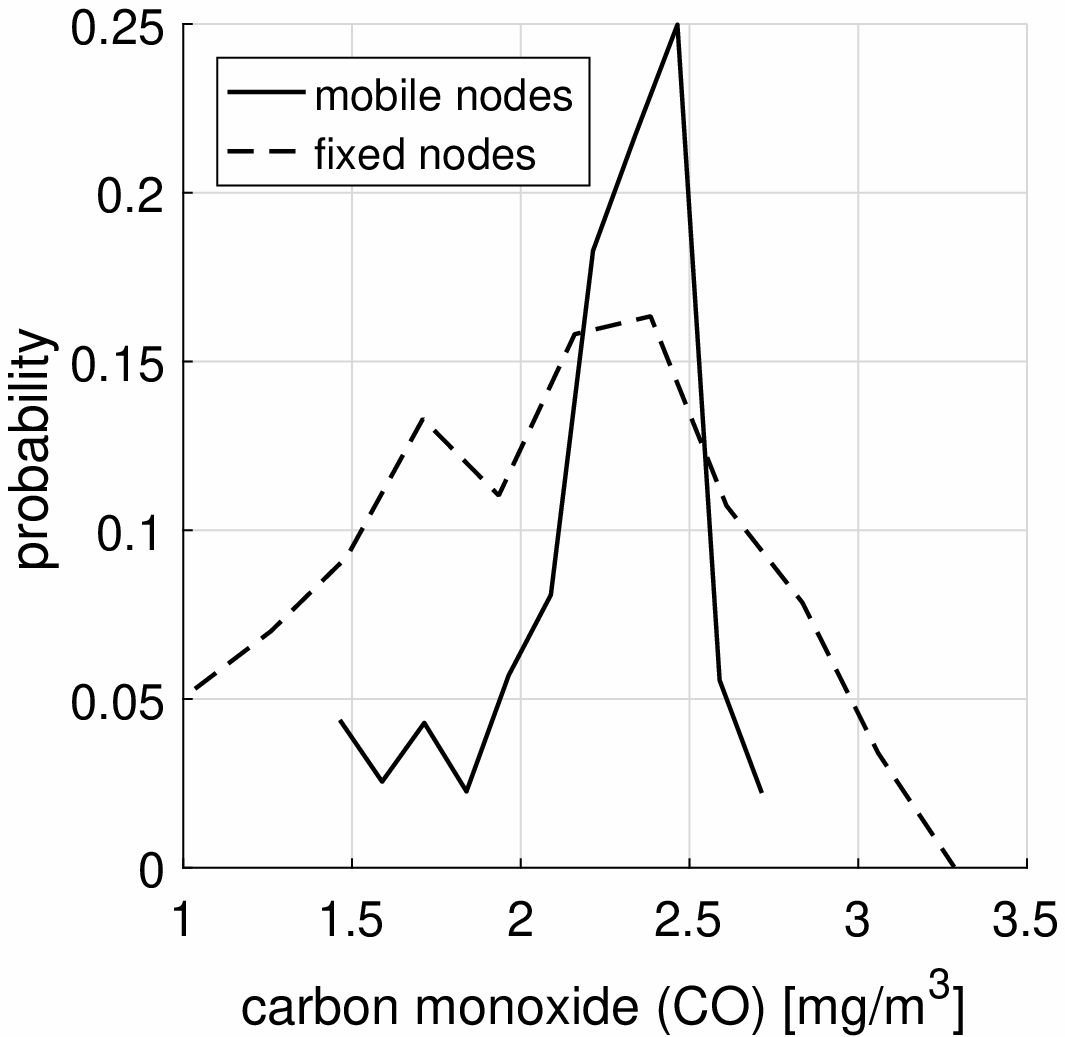}
          \label{fig:carbon_monoxide}
    	\end{subfigure}
    ~ 
        \begin{subfigure}[t]{0.22\textwidth}
          \includegraphics[scale=0.36, clip=true, trim=0 0 0 0]{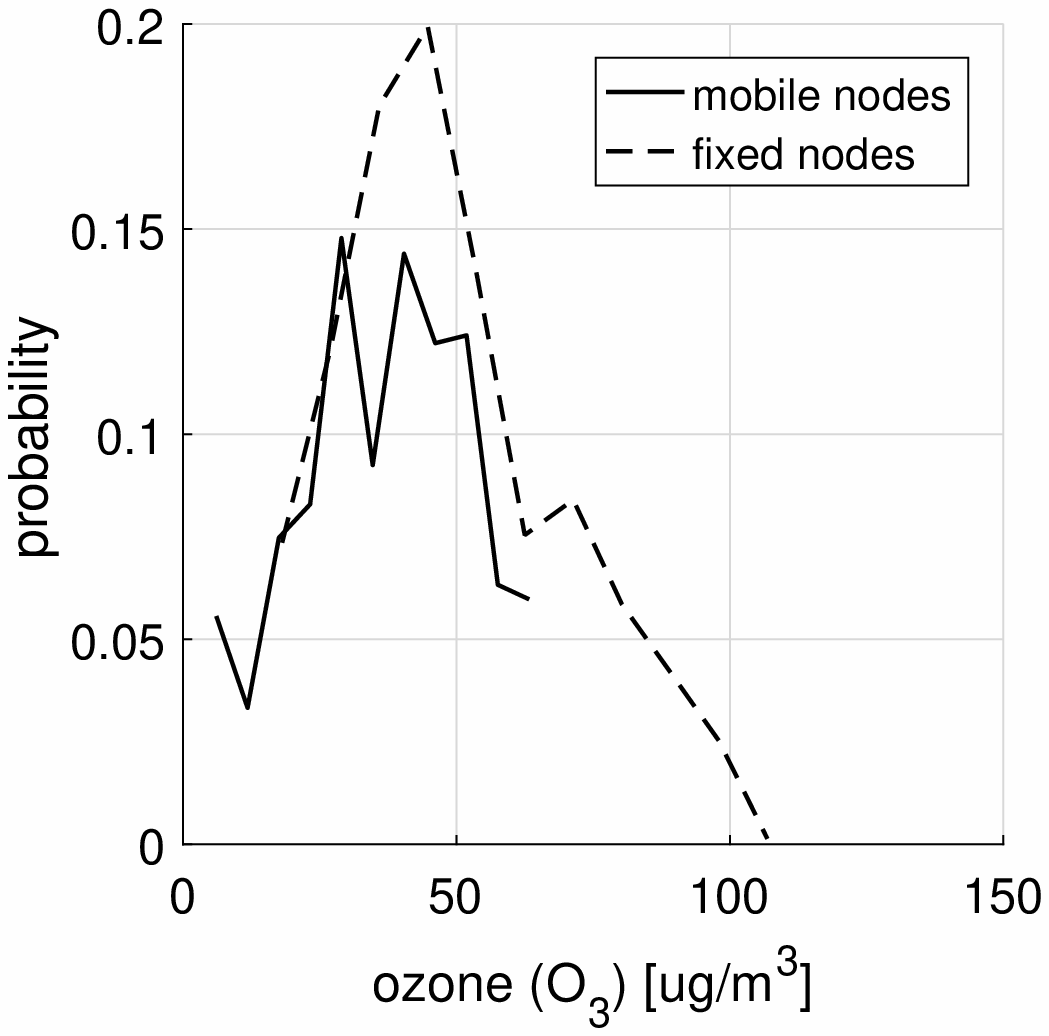}
          \label{fig:ozone}
    	\end{subfigure}
        
    \caption{Comparison among \acp{PMF} of the measurements of mobile and fixed sensor nodes}
    \label{fig:comparison}
\end{figure}

Figure \ref{fig:comparison} shows the comparison among \acp{PMF} of the pollutants and quantities collected by the mobile nodes with those collected by the closest fixed nodes\footnote{Since the large majority of the samples collected by the mobile nodes is geographically located in close proximity of two fixed stations, we considered both of them for the comparison.}; the numerical comparison (average values) is reported in Table \ref{tab:comparison}, which also provides \mbox{$\eta_{MF} = \left| 1-\dfrac{M_{mob}}{M_{fix}} \right|$}. For the same reasons as before, the \ac{PMF} of the dew point is not shown in Figure \ref{fig:comparison}.
In this case, the most significant differences are related to HC and O$_3$. It is worth highlighting here that the fixed nodes collected data at a different height w.r.t. the mobile nodes (closer to the exhaust pipes of cars), thus part of the aforementioned skew should be ascribed to that. In addition, the skew depends also on the fact that the reading of mobile nodes are collected in an area close to fixed nodes ($\approx$500m radius) and not in the same exact spatial position, thus introducing a further measurement error due to the mobility and to the corse-grained spatial precision.
\begin{table}
\begin{center}
  \resizebox{.99\hsize}{!}{
  \begin{tabular}{|c|c|c|c|} \hline
   \textbf{Measured quantity $M$} & \textbf{$M_{fix}$} & \textbf{$M_{mob}$} & \textbf{$\eta_{MF}$} \\ \hline \hline
      temperature & $\SI{14.7}{\degreeCelsius}$ & $\SI{16.9}{\degreeCelsius}$ & 0.15 \\ \hline
      relative humidity & 69.15\% & 62.05\% & 0.1 \\ \hline
      dew point & $\SI{9.9}{\degreeCelsius}$ & $\SI{9.9}{\degreeCelsius}$ & 0.001 \\ \hline
      unburned hydrocarbons (HC) & 3.08 ppmV & 5.4 ppmV & 0.76 \\ \hline 
      carbon dioxide (CO$_2$) & 424.3 ppmV & 356.9 ppmV & 0.16 \\ \hline 
      carbon monoxide (CO) & 2.06 $m$g$/m^3$ & 2.23 $m$g$/m^3$ & 0.08 \\ \hline
      ozone (O$_3$) & 49.2 $\mu$g$/m^3$ & 36.75 $\mu$g$/m^3$ & 0.25 \\ \hline
  \end{tabular}
  }
\end{center}
\caption{Comparison between the average values as reported by fixed sensor nodes ($M_{fix}$) and by mobile sensor nodes ($M_{mob}$).}
\label{tab:comparison}
\end{table}
The results presented in Tables \ref{tab:comparisonRG} and \ref{tab:comparison} indicate that the use of mobile sensor nodes for environmental monitoring is feasible (if HC is excluded), and it provides a discrete level of accuracy, with the advantage to extend the covered area in an easy and low-cost way. Furthermore, if a larger number of mobile sensors were to be deployed, the availability of large data-sets in even small geographical areas would  provide statistically reliable measurement flows. 

%% file: conclusions.tex
\section{Conclusions} 
\label{sec:conclusions}
From a technical point of view, we can conclude that the use of mobile sensor nodes can be quite effective in monitoring the air quality in cities, increasing the monitoring area with limited costs, w.r.t. to the use of only  fixed nodes, at the cost of a tolerable measurement error due to mobility and other factors, such as the height at which the mobile sensor is used. However, the real deployment of SHE system pointed out important indications on the acceptance of such a system, both by citizens and local governors. 
Specifically, the decision makers, like the municipality, demonstrated a strong interest in the collected environmental data, especially when analysed on long temporal scales (months or years). The data represent an important source of information to properly support decisions at the city level. For instance, Pisa's municipality started the definition of a new urban mobility plan in 2016, in order to try and reduce the environmental pollution due to vehicular traffic. Research scientists, including the authors, have been and are still collaborating with the local municipality on the definition of the mobility plan as technical partners, in order to design and develop innovative ICT solutions and services for the citizens (e.g., car pooling, bike sharing, smart walking paths). On the one hand, this collaboration aims at raising citizens' awareness about QoL in the city, and at further stimulating their active participation and contribution. On the other hand, it generates additional data for scientists to improve already deployed solutions and define new ones. However, to extend the deployment of the \ac{SHE} system at a city level, it would be necessary a certification process of the sensor nodes. This requires an initial investment from the city and the interest of some stakeholders, in order to manage the technological transfer of the proposed solutions by the research community. We are currently working in this direction, considering also the positive interests of citizens in this initiative.